\documentclass[twocolumn]{aastex701}

\usepackage{graphicx}
\usepackage{natbib}
\usepackage{footnote}
\usepackage{multirow}
\usepackage{amsmath}
\usepackage{amssymb}
\usepackage{rotating}
\usepackage{hyperref}
\usepackage{xspace}
\usepackage{lineno}

\defcitealias{Herrnstein05}{Herrnstein}

\newcommand{\degree}{$^{\circ}$}

\newcommand{\am}{NH$_{3}$\xspace}
\newcommand{\co}{$^{12}$CO\xspace}

\newcommand{\kms}{km s$^{-1}$\xspace}

\newcommand{\h}{$^{\mathrm{h}}$\xspace}
\newcommand{\m}{$^{\mathrm{m}}$\xspace}
\newcommand{\s}{$^{\mathrm{s}}$\xspace}

\shortauthors{Mills et al.}

\begin{document}

\title{Reconciling 3D Models for the Central 10 parsecs of the Milky Way}

\author[0000-0001-8782-1992]{Elisabeth A.C. Mills}
\affiliation{Department of Physics and Astronomy, University of Kansas, 1251 Wescoe Hall Drive, Lawrence, KS 66045, USA}
\email{eacmills@ku.edu}

\author[0000-0002-4013-6469]{Natalie O. Butterfield}
\affiliation{National Radio Astronomy Observatory, 520 Edgemont Road, Charlottesville, VA 22903, USA}
\email{}

\author[0000-0003-2300-2626]{Hauyu Baobab Liu}
\affiliation{Department of Physics, National Sun Yat-Sen University, No. 70, Lien-Hai Road, Kaohsiung City 80424, Taiwan, R.O.C.}
\affiliation{Center of Astronomy and Gravitation, National Taiwan Normal University, Taipei 116, Taiwan}
\email{}

\author[0000-0002-5776-9473]{Dani Lipman}
\affiliation{University of Connecticut, Department of Physics, 196A Hillside Road, Unit 3046 Storrs, CT 06269-3046, USA}
\email{}

\author[0000-0001-6431-9633]{Adam Ginsburg}
\affiliation{University of Florida Department of Astronomy, Bryant Space Science Center, Gainesville, FL, 32611, USA}
\email{}

\author[0000-0001-6113-6241]{Mattia C. Sormani}
\affiliation{Como Lake centre for AstroPhysics (CLAP), DiSAT, Universit\`{a} dell’Insubria, via Valleggio 11, 22100 Como, Italy}
\email{}

\author[0000-0001-9656-7682]{Jonathan D. Henshaw}
\affiliation{Astrophysics Research Institute, Liverpool John Moores University, IC2, Liverpool Science Park, 146 Brownlow Hill, Liverpool L3 5RF, UK}
\email{}

\author[0000-0002-6073-9320]{Cara D. Battersby}
\affiliation{University of Connecticut, Department of Physics, 196A Auditorium Road, Unit 3046,
Storrs, CT 06269}
\email{}

\author[0000-0003-0410-4504]{Ashley T. Barnes}
\affiliation{European Southern Observatory (ESO), Karl-Schwarzschild-Stra{\ss}e 2, 85748 Garching bei M\"{u}nchen, Germany}
\email{}

\author[0000-0001-6708-1317]{Simon C. O. Glover}
\affiliation{Universit\"{a}t Heidelberg, Zentrum f\"{u}r Astronomie, Institut f\"{u}r Theoretische Astrophysik, Albert-Ueberle-Str.\ 2, 69120 Heidelberg, Germany}
\email{}

\author[0000-0002-6379-7593]{Francisco Nogueras-Lara}
\affiliation{Instituto de Astrof\'{i}isica de Andaluc\'{i}a (IAA-CSIC), Glor\'{i}eta de laAstronom\'{i}a s/n, E-18008 Granada, Spain}
\email{}

\author[0000-0002-6753-2066]{Mark R. Morris}
\affiliation{University of California, Los Angeles, Los Angeles, Department of Physics and Astronomy, CA, 90095-1547, USA}
\email{}
\author{Juergen Ott}
\affiliation{National Radio Astronomy Observatory, PO Box O, 1003 Lopezville Road, Socorro, New Mexico 87801, USA}
\email{}

\author{Cornelia Lang}
\affiliation{Department of Physics and Astronomy, 703 Van Allen Hall, University of Iowa, Iowa City, IA 52242, USA}
\email{}

\author{Claire Cook}
\affiliation{Department of Physics and Astronomy, University of Kansas, 1251 Wescoe Hall Dr., Lawrence, KS 66045, USA}
\email{}

\author[0000-0002-5582-8521]{Xinyu Mai}
\affiliation{Department of Physics and Astronomy, University of Kansas, 1251 Wescoe Hall Drive, Lawrence, KS 66045, USA}
\email{}

\correspondingauthor{E.A.C Mills}
\email{eacmills@ku.edu}

\begin{abstract}

The construction of an accurate 3D model of the Milky Way center is necessary to understand inflow processes that drive its overall evolution, and to compare our Galactic nucleus to other galaxies' nuclei. A main point of contention is the line-of-sight location of sources observed toward the central 10 pc of the Galaxy, including recent star formation (the Sgr A East supernova remnant and Sgr A HII regions) and copious gas (the 50 and 20 km/s molecular clouds, the Circumnuclear Disk, and the Sgr A West ionized ``minispiral" that encircles the central supermassive black hole, Sgr A*). Some models place all of these structures within a radius of 5 pc from Sgr A*, while others place the 20 and 50 km/s clouds at a distance of at least 30-50 pc away from Sgr A* along the line of sight. We present new radio and millimeter observations of the molecular gas toward the central $\sim10$ pc, from which we have constructed an alternative 3D model that is consistent with both prior radio observations and orbital gas kinematics. Our model places the 20 km/s cloud, 50 km/s cloud, and Sgr A East more than 10 pc in front of Sgr A*. While this model does not conclusively rule out a connection between the 50 and 20 km/s clouds and the circumnuclear disk, we argue that prior evidence for these connections is tenuous, especially given the complex spatial and kinematic overlap of structures along the line of sight.

\end{abstract}


\section{Introduction}
\label{intro}

Our edge-on view of the Milky Way Galaxy can cause components to appear to align along the 8.3 kpc \citep{Gravity19,Gravity21} line of sight to the center. Generally, the observed properties of the central region-- large visual extinctions \citep[$A_V\sim30$, $A_{KS}\sim$2.5;][]{Nishiyama08,Schoedel10,NoguerasLara18a,NoguerasLara20b} and uniquely extreme gas conditions, e.g.\ large linewidths \citep{Shetty12} and rich organic chemistry \citep{RequenaTorres06} -- are sufficient to tell with high reliability whether or not structures are in the foreground disk of our Galaxy or located in its innermost few hundred parsecs. However, the relative placement of structures within the central R = 200 pc \citep[the central molecular zone or CMZ;][]{Morris96,Sofue22} remains contentious \citep[see, e.g., the 3D-CMZ paper series:][]{Battersby25a,Battersby25b,Walker25,Lipman25} . 

\begin{figure}[tbh]
\includegraphics[width=0.49\textwidth]{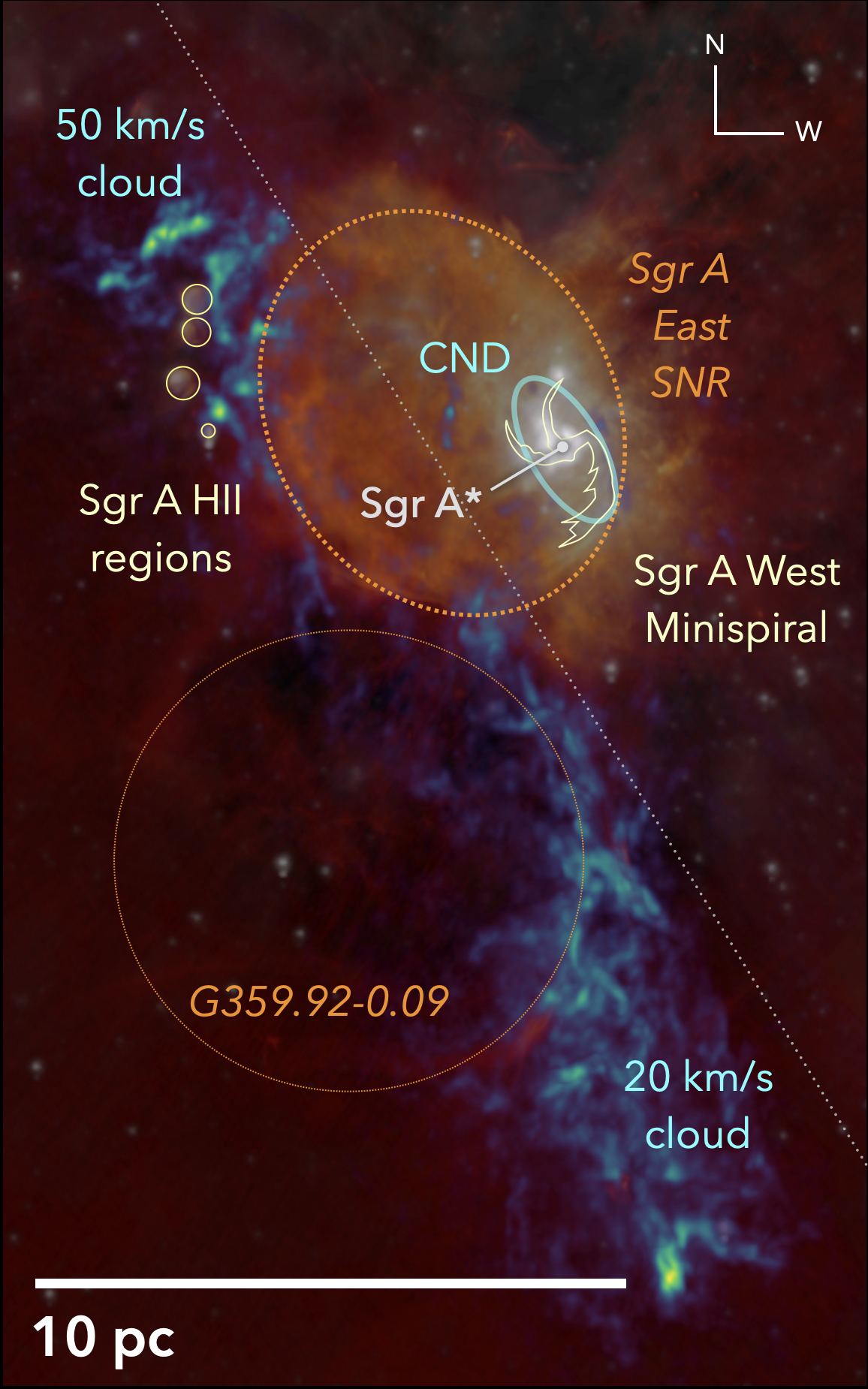}
\caption{A multiwavelength composite image highlighting structures seen toward the central 10 pc of the Milky Way. Blue/green is VLA data of emission from the (3,3) transition of \am \citep{Mills14}, highlighting the molecular gas structure in this region. Red/orange is a 6 cm VLA map from \cite{Zhao16} showing thermal (HII regions including the minispiral, outlined in light yellow solid lines) and nonthermal (supernova remnants, outlined in orange) radio continuum structures. Spitzer 8$\mu$m data \citep{Stolovy06} is shown in a grey scale. The angle of the Galactic plane is indicated with a grey dotted line at $b=-0.06$\degree.}
\label{fig:Fig-overview}
\end{figure} 

Figure \ref{fig:Fig-overview} shows a multiwavelength composite image of an approximately 10 by 20 parsec region around Sgr A*. Many different structures are visible within this region. At centimeter wavelengths, there are five main components visible in continuum emission: the \textit{Sgr A East} supernova remnant \citep[also seen in X-ray emission;][]{Ekers75,Maeda02}, a less-prominent source suggested to be a supernova remnant (\textit{G359.92-0.09}) located to the south of Sgr A East \citep[][but see also \citealt{Hsieh16} and \citealt{Zhao16} who suggest it may be an outflow feature originating from the central parsec]{Ho85,Baganoff03}, the \textit{Sgr A HII regions} \citep[A, B, C, and D;][]{Goss85,Mills11,Lau14}, the \textit{Sgr A West ``minispiral"} of ionized gas \citep{Ekers83,Zhao09,Nitschai20}, and the unresolved accretion disk of the central supermassive black hole, \textit{Sgr A*} \citep{Brown84,Melia01,EHT22}. At centimeter and millimeter wavelengths there are 3 additional components seen in molecular gas: a torus-like structure of gas and dust surrounding the minispiral \citep[the \textit{circumnuclear disk} or CND with a mass of $10^4$ M$_\odot$ and an inner radius of 1.5 pc;][]{Genzel85,Mezger96,RequenaTorres12,Lau13}, and two giant molecular clouds, \textit{M-0.02-0.07} and \textit{M-0.13-0.08}, the so-called 50 and 20 km/s clouds \citep[named for their average line of sight velocities;][]{Whiteoak74,Fukui77,Gusten80}, which appear nearby in projected distance ($R_{\rm proj}\sim$ 5-10 pc). Many of these structures are spatially extended, and so have significant overlap with the other components. 

Finally, although it is not a main focus of the analysis in this paper, the CND and minispiral are co-spatial with a nuclear stellar cluster studied at infrared wavelengths, which has a half-light radius of R $\sim$ 4 pc and a mass of $2.5\times10^7$ M$_\odot$ \citep{Schoedel14} and is centered on the supermassive black hole (see \citealt{Schoedel20} for an overview of the properties of this structure). 

In Section \ref{sec:geometry} we briefly review the existing constraints on the location of these eight main structures identified in Figure \ref{fig:Fig-overview} and outline the competing models that have been proposed for the overall 3D structure of this region. In Section \ref{data}, we describe our new observations of the molecular gas observed toward the central 10 parsecs. In Section \ref{results}, we share new findings pertaining to the physical and kinematic properties of the molecular gas, and resulting constraints on its position. In Section \ref{discussion} we further analyze these results as well as past studies of the central 10 parsecs, and present an alternative model for this region. In Section \ref{conclusions} we summarize the findings of these analyses. 

\section{Geometric Constraints from Prior Observations}
\label{sec:geometry}

\subsection{{A Review of Observations}}

Existing measurements at wavelengths from radio to X-ray have given rise to the following understanding of the central ten parsecs of the Milky Way, which we divide into seven main suppositions, describing the evidence for each in more detail. 

\subsubsection{{The Sgr A West minispiral and CND are interacting, and located in the central 1.5 parsec.}}

The minispiral consists of three streamers of ionized gas on eccentric orbits around Sgr A* at radii ranging from 0.16 to 1 pc \citep{Zhao09}. The gas is ionized by a young component of the nuclear cluster \citep{Krabbe95,Lu13}. The southernmost of the minispiral streamers is spatially and kinematically consistent with being the ionized edge of the inner CND \citep[e.g.,][]{Christopher05}, which itself may consist of multiple orbital streams of molecular gas orbiting Sgr A* at various inclinations \citep{Martin12}. While the gas kinematics of the inner CND are not consistent with a single circular orbit, warm dust observations of the material that likely encloses all of these gas streams can be well modeled as a geometric torus \citep{Lau13}. The inner radius of the CND, as traced by warm dust and dense molecular species, is 1.5 pc \citep{Christopher05,MonteroCastano09,Lau13}, while gas traced by CO and streamers of CS extends out to radii of at least 5 pc \citep{Oka11,Liu12}, with atomic carbon emission extending up to 7 pc to the south \citep{Serabyn94,Morris96}. 

\subsubsection{{Sgr A West lies in front of Sgr A East.}}
\label{sec:ii}
\cite{YusefZadeh87} first observed the Sgr A West minispiral in absorption at 90 cm, superposed against emission from the Sgr A East supernova remnant (SNR). This was also independently observed by \cite{Pedlar89}, who conclude that free-free absorption of the synchrotron photons from the Sgr A East SNR is the cause of the dip in emission observed at the position of Sgr A West (their figure 7). Absorption against Sgr A East can only occur if this minispiral lies in front of the SNR along our line of sight. 

\subsubsection{{The 50 km/s cloud is interacting with Sgr A East.}}
\label{sec:iii}
Initial arguments for this interaction were based primarily on the striking morphology of the 50 km/s cloud, which parallels the northeastern edge of the SNR shell \citep{Genzel90,Serabyn92}. Stronger evidence for interaction comes from the detection of OH 1720 MHz masers on the edge of Sgr A East at velocities consistent with the gas in the 50 km/s cloud \citep{YusefZadeh96,Karlsson03,Sjouwerman08}. This class of maser appears to be exclusively produced in an X-ray irradiated shock environment where a supernova remnant impacts a nearby cloud \citep{Frail96,YusefZadeh03}.  The gamma ray spectrum of Sgr A East is also consistent with the interaction of a SNR with a dense molecular cloud, and with Sgr A East being a source of cosmic ray acceleration \citep{Fatuzzo03,Grasso05,Cavasinni06}. Further evidence for interaction comes from the chemistry of the gas, which shows signatures of being both highly irradiated and strongly shocked. The 50 km/s cloud gas is notably bright in 
CI \citep{Martin04, Tanaka11, Garcia16, Tanaka21}, particularly when compared to the rest of the CMZ  \citep[][ their Figure 1]{Tanaka11}, which is attributed to a high level of cosmic ray irradiation from the Sgr A East SNR. \cite{Lee08} also see shock-excited H$_2$ emission in the 50 km/s cloud around the periphery of the Sgr A East shell.   

\subsubsection{{Sgr A West and the CND may be physically interacting with Sgr A East.}}
\label{sec:iv}
An interaction between Sgr A West, the CND, and Sgr A East has been suggested due to the morphological similarity of diffuse X-ray emission and the minispiral western arm, where Sgr A* and Sgr A West are interpreted to be enveloped by the shell of Sgr A East \citep{Baganoff03}. Studies have also argued that a `ridge' feature seen in X-ray emission to the northeast of Sgr A* is a shock front where the winds of stars in the young nuclear cluster encounter supernova ejecta from Sgr A East \citep{Maeda02,Rockefeller05,Zhang23}. \cite{Zhao16} further argue that a radio feature within the Sgr A East SNR may be the result of the reflection of the supernova shockwave off of the CND. 
Sgr A East has also been suggested to have engulfed part of the CND based on the presence of OH and H$_2$ shock tracers in the CND gas, which they also suggest is connected to the western edge of Sgr A East  \citep{YusefZadeh99}. However, this interpretation is disputed by \cite{Sjouwerman08}, who do not find indications of direct interaction between Sgr A East and the CND in their OH absorption data, and further argue that given the physical conditions in the CND, the existence of OH masers do not require a supernova remnant shockwave to form.

\subsubsection{{The Sgr A HII regions A-D are embedded in the 50 km/s cloud.}}
Radio and infrared recombination line observations \citep{Goss85,YusefZadeh10} show that this group of 4 HII regions that roughly parallel the outer edge of Sgr A East have velocities consistent with the molecular gas in the 50 km/s cloud. \cite{Mills11} measured extinctions to the HII regions by comparing thermal radio continuum to Paschen alpha emission, finding extinctions of 3.7 magnitudes at 1.87 $\mu$m  \citep[$A_V \sim 45$, adopting the extinction law of ][]{Nishiyama08} for 3 of the sources, suggesting that most of them lie on the front (near side) edge of the cloud. However, the southernmost source (D) is consistent with being more embedded within the cloud ($A_{1.87}=5.9$, $A_V = 71$). Though their distribution along the edge of the SNR makes it tempting to ascribe their formation to a density enhancement due to the shock impact of Sgr A East, the timescales are fundamentally incompatible: ultracompact HII regions have ages of $10^4-10^5$ years \citep{Churchwell02}, while the age of Sgr A East is estimated to be less than 1500-1700 years from hydrodynamical simulations \citep{Fryer06,Zhang23}, although some models and interpretations prefer an older age of 9,000 to 10,000 years \citep{Zhao13,Ehlerova22}. 

\subsubsection{{The 50 km/s cloud is connected to the 20 km/s cloud.}} 
This connection has not always been accepted: \cite{Gusten80} originally argued against it because the velocity gradients of the two clouds are different, and they did not appear to continuously connect in early H$_2$CO absorption studies. However, more recent studies \citep[e.g.][]{Genzel90,Ho91,Herrnstein05} identified a 
`Molecular Ridge' of connecting material between the two clouds. A continuous structure in both position and velocity space is seen even more clearly in large scale data sets of the entire CMZ. \cite{Kruijssen15} place the 50 and 20 km/s cloud on part of an even larger feature, ``Stream 1", spanning 100 pc. A robust analysis of CMZ position-position-velocity data by \cite{Henshaw16} also confirms that the two clouds are connected smoothly and coherently by a velocity gradient of $\sim$2.3 km s$^{-1}$ pc$^{-1}$ over a projected distance of $\sim$17 pc. This is also consistent with the recent analysis of \citep{NoguerasLara26} who find a similar signature from both clouds using a combination of extinction and stellar kinematics and densities. Bucking this consensus however, a recent model by \cite{Sofue25} suggests that these clouds may not be connected, and could also belong to separate `spiral arms' separated by 30 pc along the line of sight. 

\subsubsection{{The 20 km/s cloud may or may not be connected to the CND.}}
Initial claims of a connection were based on the detection of a velocity gradient along a structure at the northern tip of the 20 km/s cloud (the `southern streamer') which appears to adjoin the CND \citep{Okumura91}. Other early studies reinforced the idea of a link based on the proximity of the two structures \citep{Zylka90,Dent93,Marshall95} and on evidence of heating and increases in line widths as the southern streamer approaches the CND \citep{Coil99,Coil00}. However, a later study argued that the southern streamer does not connect to the CND, because it has neither the velocity gradient nor curved morphology that a cloud on a more plunging orbit toward the nucleus should have \citep{Herrnstein05}. 
Separately, \cite{Takekawa17} proposed another connection between the 20 km/s cloud and a negative-longitude extension of the CND, though \cite{Hsieh17}) dispute this interpretation of a kinematic connection, finding that it does not fit within their kinematic model.

\subsection{A Review of Models}

\begin{figure*}[tbh]
\includegraphics[width=1.0\textwidth]{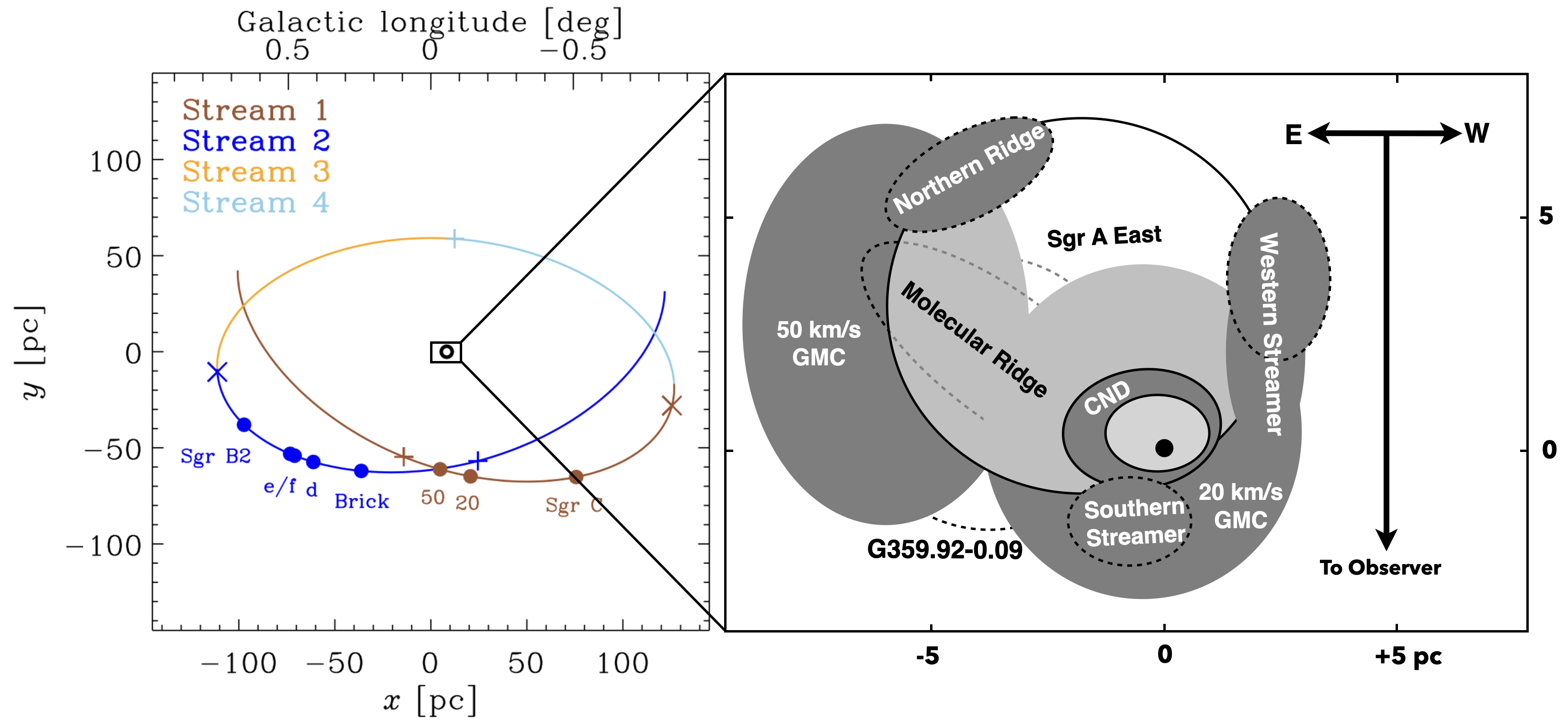}
\caption{Left: A reproduction of the top-down model of \cite{Kruijssen15}, showing the location of the 50 and 20 km/s clouds on orbital streams that place them $>$50 pc in front of Sgr A* along the line of sight. The inner $\sim$ 10 parsec region from the \citepalias{Herrnstein05} model is indicated as a box around Sgr A*. Right: A recreation of the model schematic from \cite{Lee08}, showing a top-down view of the arrangement of sources first suggested by \cite{Herrnstein05}. As a scale bar was not originally included in this model, we scale this model to the observed major axis of Sgr A East (8.3 pc).}
\label{fig:Fig-old-model}
\end{figure*}

\subsubsection{Models for the inner 10 parsecs}

Models of the central $R\lesssim 10$ parsecs have been primarily informed by radio-frequency observations. Initial models, as now, sought to explain the relative placement of molecular gas, ionized gas, and nonthermal plasma relative to the central supermassive black hole. 
Early models disagreed widely about whether the 50 and 20 km/s clouds were close to Sgr A*, and whether they were in front or behind \citep[e.g.,][]{Whiteoak74,Gusten80,Zylka90,Genzel90,Okumura89,Coil99}

The currently most widely-accepted model was put forward by \citet{Coil00}, based on VLA observations of \am (1,1) and (2,2). It was refined by \cite{Herrnstein05} using additional observations of the \am (3,3) and (6,6) lines. In this model, shown in the second panel of Figure \ref{fig:Fig-old-model}, Sgr A West lies on the front side of (or slightly within) the Sgr A East SNR. The 50 and 20 km/s clouds are connected to each other and both interact with Sgr A East (the latter through the `Western Streamer' which is assumed to originate from the 20 km/s cloud and to lie along the western extent of the Sgr A East shell). The Sgr A East SNR is also connected directly to the CND by the `Northern Ridge', as initially argued in \cite{McGary01}. The `Southern Streamer' extension of the 20 km/s cloud lies in front of and close to the CND, but is not required to connect directly to it. Unlike some previous models, the kinematics of the 50 and 20 km/s clouds are not explained in the context of the overall orbits of other giant molecular clouds observed in the Galactic center. There have been slight adjustments to this model \citep[e.g., suggesting the `Western Streamer' is an extension of the 20 km/s cloud, that the `Southern Streamer' is directly connected to the CND, or that the `Northern Ridge' is not connected directly to the CND; ][]{Lee08,Ferriere12}. However, these are relatively minor adjustments, as in these models all of these features are still shown as existing in close ($<5$ pc) spatial proximity to each other. In this sense, this model has remained largely unchanged since it was first proposed, and we subsequently refer to it in this paper as the `\citetalias{Herrnstein05} model'. The defining characteristic of the \citetalias{Herrnstein05} model is then this assumption that all of the prominent structures in this region observed at radio wavelengths in continuum and molecular lines lie within a radius of 5 pc from Sgr A*, and are truly as close to the center as they appear to lie in projection.

\subsubsection{Models for the entire CMZ}

In contrast to models of just the central 10 pc, models of the geometry of the entire CMZ (R$\sim$150 pc) are primarily based on fits to large-scale gas kinematics with some qualitative constraints from absorption (e.g., the appearance of sources as infrared dark clouds). As such, these models tend not to have the complex dependencies of models of the inner 10 parsecs, which attempt to reconcile many different constraints from dissimilar observations. Models generally agree that the dense gas is not arranged in a filled disk but instead is arranged in orbital structures at distinct radii. However, they otherwise disagree on the shape, size, and orbital velocity of these structures. 

Proposed models include elliptical `x2' orbits \citep[][Lipman et al. in prep.]{Binney91,Sormani15,Tress20,Walker25}, closed loops with constant orbital velocities \citep{Molinari11}, spiral arms \citep{Sofue95,Sawada04,Ridley17,Sofue25}, and open or precessing orbits \citep{Kruijssen15}. Orbital properties of some of these main models are summarized in Table \ref{tab:models}. The placement of orbits relative to Sgr A* is a major contention between these models. The initial spiral arm model of \cite{Sofue95} places the 50 and 20 km/s clouds on the far side of the CMZ, while revised spiral arm models from \cite{Ridley17} and \cite{Sofue25} place both clouds from 8 to 50-100 pc in front of Sgr A*. \cite{Molinari11}, place the 50 and 20 km/s clouds in front of Sgr A* and within 20 pc of the center. \cite{Walker25} present a modification of this model that provides an improved fit to observed gas kinematics, {but their orbital model has gas at the position and velocity of the} 50 and 20 km/s clouds on the far side of the CMZ, {while their absorption measurements suggest instead that these clouds are located on the near side.} While \cite{Kruijssen15} place the 50 and 20 km/s clouds in front of Sgr A*, they have them at larger distances than \cite{Molinari11}: roughly 60 pc in front of Sgr A*. \cite{Tress20} use examples from their simulations to suggest that the 50 and 20 km/s clouds could be located off of the main CMZ gas orbit, lying somewhere between the CND and other CMZ clouds, at $r<30$ pc. Lipman et al. (submitted) also perform orbital fitting informed by a statistical analysis of published positional constraints. They find the CMZ (not including the 20 and 50 km/s clouds) to be well fit by a range of nested x2 orbits with major axis extents between 60 - {140} pc. They suggest that the kinematics of the 50 and 20 km/s clouds have additional complexity not described by this average orbital model, and constrain the positions of {the 50 and 20 km/s} clouds to be at a distance of {31$\pm$14 pc and 43$\pm$19 pc} in front of Sgr A*, {respectively}. For all of these models however, it is important to note that unless both the 50 and 20 km/s clouds are within a radius of 5 pc, they are not consistent with the constraints adopted by the \citetalias{Herrnstein05} model. 

\cite{Lipman25} and \cite{Walker25} assess these 3D models using dust extinction and H$_2$CO absorption, finding that the \cite{Kruijssen15} model is more consistent with line-of-sight distance constraints from these observations than either the `spiral arm' model or the `twisted ring' model. Their {position constraints (independent of separate kinematic modeling)} place both the 50 and 20 km/s clouds as likely to be on the near side of the CMZ, though both papers note additional uncertainties for these clouds due to their proximity to bright infrared and radio emission from the Milky Way nucleus. This is also consistent with a recent analysis of stellar kinematics, based on attenuation by both clouds, which places the 50 {and 20} km/s clouds on the front side of the CMZ, at {distances of 43 $\pm$ 8 pc and 56 $\pm$ 11 pc} from Sgr A*, {respectively} \citep{NoguerasLara26}.

\begin{table*}[tbh]
\centering
\caption{Orbital Models for the CMZ}
\begin{tabular}{llllll}
\hline\hline
Model & Shape & Orbital & Semimajor  & Distance from Sgr A* & References \\
 & & Speed & Axis & of 50 and 20 km/s clouds & \\
 &  & (km/s) & (pc) & (pc) & \\
\hline
Spiral Arms & 2 Arms & 150 &120 & $\sim$ 100 (behind) & \cite{Sofue95}\\
 & 6 Arms & 150 & 2.3-120 & $\sim$ 8 and $\sim$ 20 (in front) & \cite{Sofue25}\\
Twisted Ring & closed loop & 80  & 60 $\times$ 40 & $\sim$ 20 (in front) & \cite{Molinari11} \\
 x2 & elliptical & 130 & 90 $\times$ 55 & $\sim$ 50 (behind) & \cite{Walker25} \\
& toroidal  & {110-130} & {30-70} &  {31$\pm$14 and 43$\pm$19} pc (in front) & Lipman et al. (submitted) \\
Open Stream & rosette & 100-200 & 50-100 & $\sim$ 50  (in front) & \cite{Kruijssen15} \\
\hline\hline
\end{tabular}
\label{tab:models}
\end{table*}

\subsection{Tensions between models and observations}

The primary tension between kinematic models of the entire CMZ and the \citetalias{Herrnstein05} model is the location of the 50 and 20 km/s clouds. If the kinematic models of the entire CMZ are correct and the 50 km/s cloud is even as much as 10 pc away from Sgr A*, this is inconsistent with the requirement that Sgr A East (with an apparent radius of 4 pc) is interacting both with the 50 km/s cloud and Sgr A*. Alternatively, if the \citetalias{Herrnstein05} model is correct and the 50 and 20 km/s clouds are as close to Sgr A* as they appear to lie in projection (at radii of $\sim$5 pc), \cite{Kruijssen15} show that the radial velocity gradient of these clouds should be significantly steeper. While this is the major disagreement, there are additional observational constraints that call the \citetalias{Herrnstein05} model into question:

\begin{itemize}
\item If the 50 km/s cloud is behind Sgr A*, why does it appear as an infrared dark cloud in front of the nuclear star cluster \citep{Schoedel14}? 
\item If the `East CND' and `Southern Streamer' (see Panel 2 of Figure \ref{fig:Fig-old-model}) are at the same distance from Sgr A* as the CND, why are they more prominent in absorption than the CND in Spitzer maps \citep{Stolovy06}?
\item If Sgr A West and the surrounding CND are in front of Sgr A East, why do the X-ray observations of Sgr A East not show an absorption feature from the CND due to Compton scattering from the column of H and H$_2$ \citep{Baganoff03,Ponti15}?
\item If the `Southern Streamer' is truly as close to the CND as it appears in projection, why is its edge not ionized \citep{Zhao16}, and why is it chemically distinct from the CND \citep[e.g., having an apparently higher abundance of the easily photodissociated molecule H$_2$CO;][]{Martin12}.
\end{itemize}

This paper seeks to resolve the tension between models of the central 10 parsecs and the entire CMZ by presenting an alternative model of the inner 10 parsecs, replacing the \citetalias{Herrnstein05} model with one that attempts to reconcile observational constraints and larger-scale  models of the CMZ gas. 

\section{Data}
\label{data}
Our analysis focuses primarily on new datasets which represent a substantial improvement over the observations on which many of the existing 3D models of the central 10 pc have been based.

\subsection{Radio Data}
The observations presented in this paper include centimeter-wave spectral line data collected with the Karl G. Jansky Very Large Array (VLA), a facility of the National Radio Astronomy Observatory\footnote{The National Radio Astronomy Observatory is a facility of the National Science Foundation operated under cooperative agreement by Associated Universities, Inc.}, and the Robert C. Byrd Green Bank Telescope (GBT).

\subsubsection{VLA observations}

The VLA observations were made on January 8, 2012 as part of a larger survey that covers the Sgr A complex as well as several other Galactic center clouds. These observations are described further in separate papers \citep{Mills15,Ludovici16,Butterfield18,Mills18b,Butterfield22}. The calibration of these data is the same as described in \cite{Mills15}, and additional details of this procedure can be found there. In this paper we focus on GCM-0.02-0.07 and GCM-0.13-0.08 (the  `50 km/s' and `20 km/s' clouds, respectively) as well as the circumnuclear disk (`CND') surrounding the central supermassive black hole, Sgr A*. All three clouds were observed as part of a seven-pointing mosaic covering an area of 4.5$'$ by 9$'$ (11$\times$22 pc), as shown in Figure \ref{fig:Fig-velocity}. The data were taken in the hybrid DnC array configuration of the VLA, a configuration that yields a nearly circular beam shape given the low altitude of the Galactic center from the VLA site. The resulting angular resolution is $\sim3''$, which corresponds to a spatial resolution of $\sim$0.1 pc at the assumed Galactocentric distance of 8.3 kpc \citep{Gravity21}. 

Observations were made of six metastable rotational transitions of ammonia (\am) toward these two clouds: the (1,1) line at 23.69450 GHz, the (2,2) line at 23.72263 GHz, the (3,3) line at 23.87013 GHz, the (4,4) line at 24.13942 GHz, the (5,5) line at 24.53299 GHz, and the (6,6) line at 25.05603 GHz. All \am lines were observed simultaneously with the K-band receiver, using the WIDAR correlator in a setup covering the frequency ranges 23.5-24.5 and 25 - 26 GHz. The spectral resolution was 250 kHz (2.7 - 3.2 \kms) for all lines but the \am\, (1,1) and (2,2) lines, for which the spectral resolution was 125 kHz (1.6 \kms) in order to resolve their hyperfine structure. The total integration time for each pointing was $\sim$30 minutes.  In this paper, we focus on observations of the (3,3) line, as it is the brightest of these six transitions, and has less prominent hyperfine structure than the (1,1) and (2,2) transitions. 

\subsubsection{GBT observations}

In order to increase sensitivity to spatially-extended structure, the VLA \am observations were combined with GBT observations made on November 7, 2011. The GBT data consist of a dual-polarization on-the-fly mosaic using four beams of the K-band Focal Plane Array \citep[KFPA;][]{Masters11} covering a $10'\times12'$ field in Right Ascension and Declination that covers the entire area mapped with the VLA. The effective integration time per beam per polarization was $\sim$ 12 s. The spatial resolution of the GBT data at the frequency of the \am\, lines is $\sim31''$ (1.2 pc). The (1,1) through (6,6) \am\, lines were observed simultaneously with the VEGAS spectrometer \citep{Prestage15} in two spectral bands having a 50 MHz total bandwidth, centered on 23.870129 GHz and 25.056025 GHz. The spectral resolution was 12.207 kHz ($\sim$0.15 \kms). An analysis of the \am\, (3,3) and (6,6) line data was previously published in \cite{Minh13}, and additional details of the data calibration can be found there. 

Combination of the GBT and VLA data was performed using a procedure that is similar to that used in \citet{Liu13} (see their Appendix A) to combine the SMA 0.86 mm and JCMT SCUBA continuum observations towards the central 10 pc region around Sgr\,A*.
The procedure was based on the {\sc Miriad} software package, which has now been re-organized as a comprehensive tool, ALMICA\footnote{https://github.com/baobabyoo/almica}.

In preparation for combination, the GBT data were spectrally smoothed and regridded to the velocity resolution of the VLA data (250 kHz or 3.17 \kms). The spectrally smoothed and regridded GBT image cubes were then deconvolved using the {\sc Miriad}-{\tt clean} task, and then were implemented with the primary beam attenuation of the VLA mosaic using the {\sc Miriad}-{\tt demos} task. 
Afterwards, we used the {\sc Miriad}-{\tt uvrandom} and {\tt uvmodel} tasks to convert the deconvolved and VLA-primary-beam attenuated GBT image cubes to visibilities (hereafter the GBT visibilities) that were sampled over the {\it uv} distance range of 0--3.5 $k\lambda$.
We then jointly imaged the GBT and VLA visibilities using the {\sc Miriad}-{\tt invert} and {\tt mossdi} tasks. 
In the joint imaging, the relative weights between the GBT and VLA visibilities were manually optimized by artificially resetting the system temperature columns, until we can visually see the contributions from both the GBT and VLA observations in the dirty images (i.e., the inverse-Fourier transformed visibilities) while the synthesized beams of the joint images are approximately Gaussian.

The deconvolution algorithms used throughout this procedure do not conserve integrated flux densities.
In most cases, this led to underestimates of extended emission even after the combination of the single-dish data (c.f. \citealt{Monsch18}).
To mitigate this issue, in the end, we used the {\sc Miriad}-{\tt immerge} task to linearly combine the GBT+VLA joint image cubes we produced and the spectrally smoothed and regridded GBT image cubes. 
There are still $\lesssim$1\% of systematic intensities errors due to the fact that the {\sc Miriad}-{\tt immerge} task does not preserve the Gaussianity of the synthesized beam (see Figure 7 of \citealt{Jiao22})\footnote{The CASA-{\it feather} task is subject to a similar but worse issue.}, which is negligible as compared to the thermal noises in our present study. 
The resulting \am (3,3) image has a beam size of $2.9''\times3.7''$ and a per-channel noise of 1.3 mJy/beam (0.26 K).

\subsection {ALMA data}
In addition to the larger VLA mosaic, we have both pointings and small mosaics observed toward the CND at millimeter wavelengths with the Atacama Large Millimeter/submillimeter Array (ALMA). {Lines observed in these datasets are listed in  in Table \ref{tab:ALMA}}.

\subsubsection{Band 3 CO pointing toward Sgr A*}

The data analyzed here are part of a larger survey that was observed using ALMA in Cycle 7 (Project code 2019.1.01240.S; PI: E.A.C. Mills). The observations were completed over eight sessions, with an average of 45 antennas, between 2019 October 31 and November 24. This survey consisted of 25 pointings within the central 5$\degr$ of the Galaxy. In this paper, we present results from a single pointing with a field of view (half-power beamwidth) of 52$''$ (2 pc) toward Sgr A*, centered at $\alpha=17^h45^m40^s$, $\delta=-29\degr00'28''$. The location of this pointing is shown in the left panel of Figure \ref{fig:Fig-velocity}. The observations were made in the C43-2 configuration, with baselines ranging between 15 and 697 m.

The observations were made using two frequency settings, both at 3 mm (ALMA Band 3), over a total of eight spectral windows. Five of these had bandwidths of 234 MHz ($\sim$640 \kms) and resolutions of 141 kHz (0.37 \kms) and targeted multiple carbon monoxide (CO) isotopologues as well as the H40 $\alpha$ recombination line, while the remaining three had bandwidths of 1875 MHz (5500 \kms) and resolutions of 1.3 MHz (3.5 \kms), covering continuum emission. Observations from this survey toward Sgr E have been published in \cite{Wallace22}.

The calibration of the data was performed in the Common Astronomy Software Applications package (CASA versions 5.6.1–8) with the ALMA pipeline.  The imaging of both line and continuum emission was performed in CASA, using the {\tt tclean} task, with a robust weighting of 1.0, so that the observations were sensitive to a combination of both point source emission and extended emission. 

This paper focuses its analysis just on the $^{12}$CO $J=1-0$ transition, which has a rest frequency of $115.2712018$ GHz. The imaged cube for this line has 0.28$''$ pixels (0.011 pc) and a beam size of $2.4'' \times 1.4''$ or $ 0.09 \times 0.06$ pc. It is sensitive to emission on size scales up to 21$''$ (0.84 pc). The per-channel noise in this cube is 13.3 mJy/beam (0.395 K).   

{\subsubsection{Band 3 and 6 CND Mosaic}}

\begin{table*}[tbh]
\caption{ALMA observations of the CND}
\begin{tabular}{llllllll}
\hline\hline
 Molecule & Transition  & E$_\mathrm{upper}$ & Critical& Rest & Angular  & Velocity & Per-Channel \\
   &  & & Density\footnote{At a temperature of 60 K} & Frequency &Resolution & Resolution &Line Sensitivity  \\
    &  & K & cm$^{-3}$ & GHz &($''$) & \kms & K  \\
\hline
HC$_3$N & $10-9$ & 24.0 & $1.2\times10^6$ & 90.97902 & 1.95 $\times$ 1.39  & 3.4 & 0.23 \\
N$_2$H$^+$ & $1-0$ & 4.5 & $1.9\times10^5$ & 93.17340 & 1.90 $\times$ 1.35  & 3.4 & 0.27 \\
SO & $6_6-5_5$ &56.5 & $3.8\times10^6$ & 258.25583 & 1.42 $\times$ 0.97  & 1.2 & 0.10 \\
H$^{13}$CN & $3-2$ & 24.9 & $7.0\times10^7$ & 259.01179 & 1.42 $\times$ 0.96 & 1.2 & 0.12 \\
CH$_3$OH & $5_{2,3}-4_{1,3}$ &57.1 &$5.2\times10^7$ & 266.83813 & 1.23$ \times$ 0.92  & 1.2 & 0.14  \\
HNC & $3-2$ & 26.1 & $8.5\times10^6$ & 271.98111 & 1.23 $\times$ 0.88 & 1.2 & 0.15 \\
\hline\hline
\end{tabular}
\label{tab:ALMA}
\end{table*}

We also analyze a data set from ALMA in Cycle 2 (project code 2013.1.00857.S; PI: E.A.C. Mills) which made observations towards the CND in both Band 3 and Band 6. 

The Band 3 observations analyzed here consist of an 11-pointing mosaic covering a $75''\times100''$ (3$\times$4 pc) field of view. The location and size of this mosaic is shown in the left panel of Figure \ref{fig:Fig-velocity}. Observations were conducted on July 21, 2014 and December 6, 2014. Observations were made in a single frequency setting covering four spectral windows centered at 90.7 GHz, 92.6 GHz, 102.75 GHz, and 104.6 GHz. All spectral windows were configured with bandwidths of 1.875 GHz and channel widths of 1.13 MHz (3.2 - 3.7 \kms). 

The Band 6 observations analyzed here consist of a 68-pointing mosaic covering a $75''\times100''$ (3$\times$4 pc) field of view. Observations were conducted on April 6, 2015. Observations were made at two frequency settings, covering eight spectral windows centered at 252.55 GHz, 254.72 GHz, 256.7 GHz, 258.7 GHz, 266.72 GHz, 269.70 GHz, 271.5 GHz, and 273.5 GHz. All spectral windows were configured with bandwidths of 1.875 GHz and channel widths of 1.13 MHz (1.2 - 1.3 \kms). 

The Band 3 observations consisted of two configurations of the 12m array (C34-2/1 with minimum baselines of 14m and maximum baselines of 348 m and C34-4/5 with minimum baselines of 17 m and maximum baselines of 783 m), while the Band 6 observations used only a single configuration of the 12m array (C34-1, with minimum baselines of 14 m and maximum baselines of 327 m). The 12m observations were combined with observations from the 7m array (ALMA Compact Array) and total power dishes in order to recover emission at all spatial scales. The data were calibrated manually by North American Science Center. The 7m and 12m data were then jointly imaged using CASA, and the images were then feathered with the total power data. Properties of the resulting images are given in Table \ref{tab:ALMA}. 

\section{Results}
\label{results}

\subsection{A New \texorpdfstring{\am}{Ammonia} map of the central 10 parsecs}
\label{sec:Ammonia}

\begin{figure*}[tbh]
\includegraphics[width=0.99\textwidth]{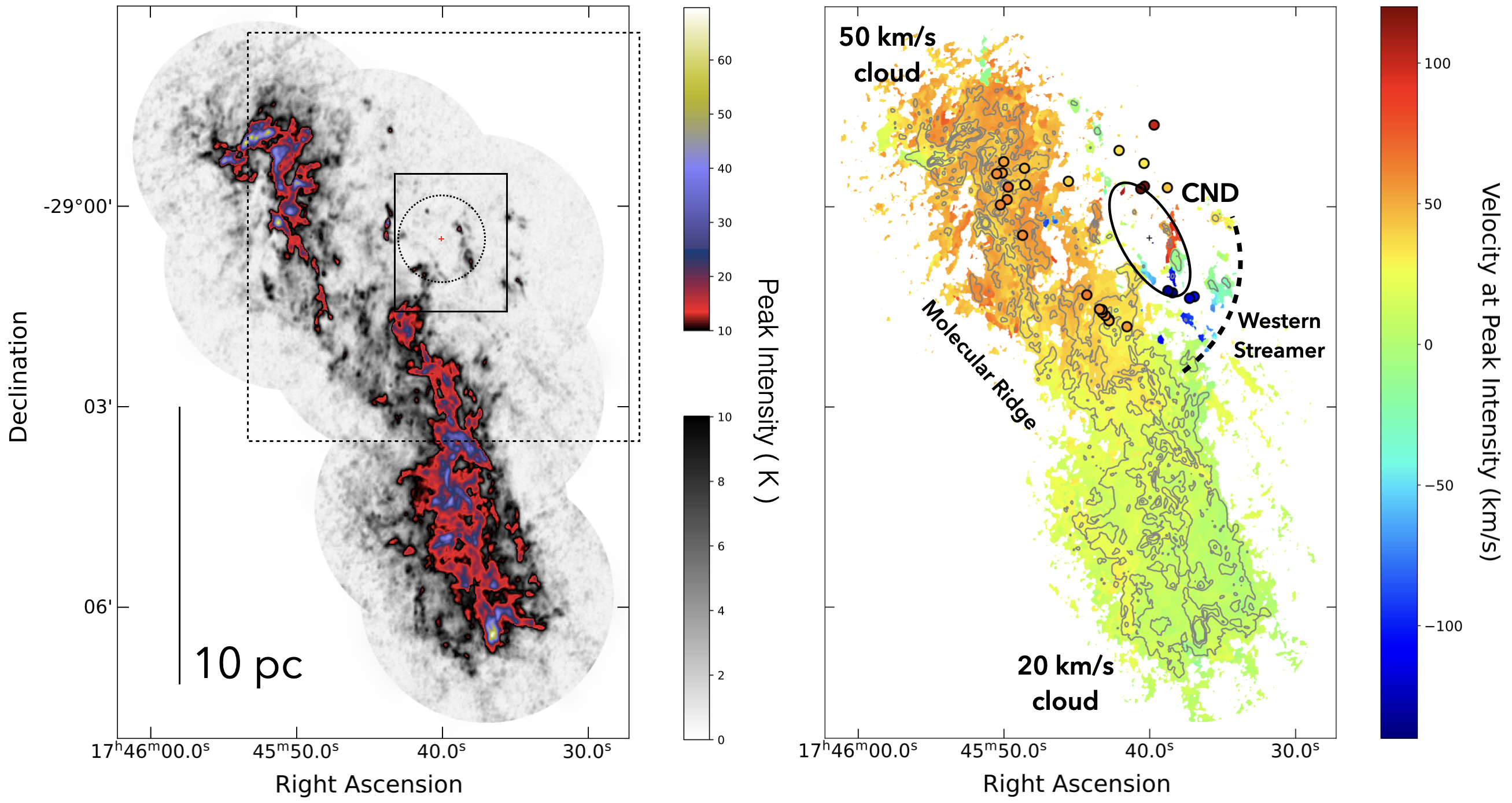}
\caption{Left: An integrated intensity map of the \am (3,3) line. The location of Sgr A* is shown as a red cross. Overlaid on this map are the fields of view of other data sets discussed in this paper. The field of view of the ALMA \co measurements is shown as a dotted circle. The field of view of the ALMA CND mosaics is shown as a solid rectangle. The field of view of the \cite{Herrnstein05} \am mosaic from Figure \ref{fig:Fig-old-model} is shown as a dashed square. Right: A map of the velocity in the \am (3,3) cube at the location of the peak intensity for each pixel. The locations of the CND and the `Molecular Ridge' connecting the 50 and 20 km/s clouds are indicated. Overlaid on the map are the 1720 MHz OH masers from \cite{Sjouwerman08}. The velocities of these masers are displayed with the same color bar as the background image}
\label{fig:Fig-velocity}
\end{figure*} 

\begin{figure*}[tbh]
\includegraphics[width=0.36\textwidth]{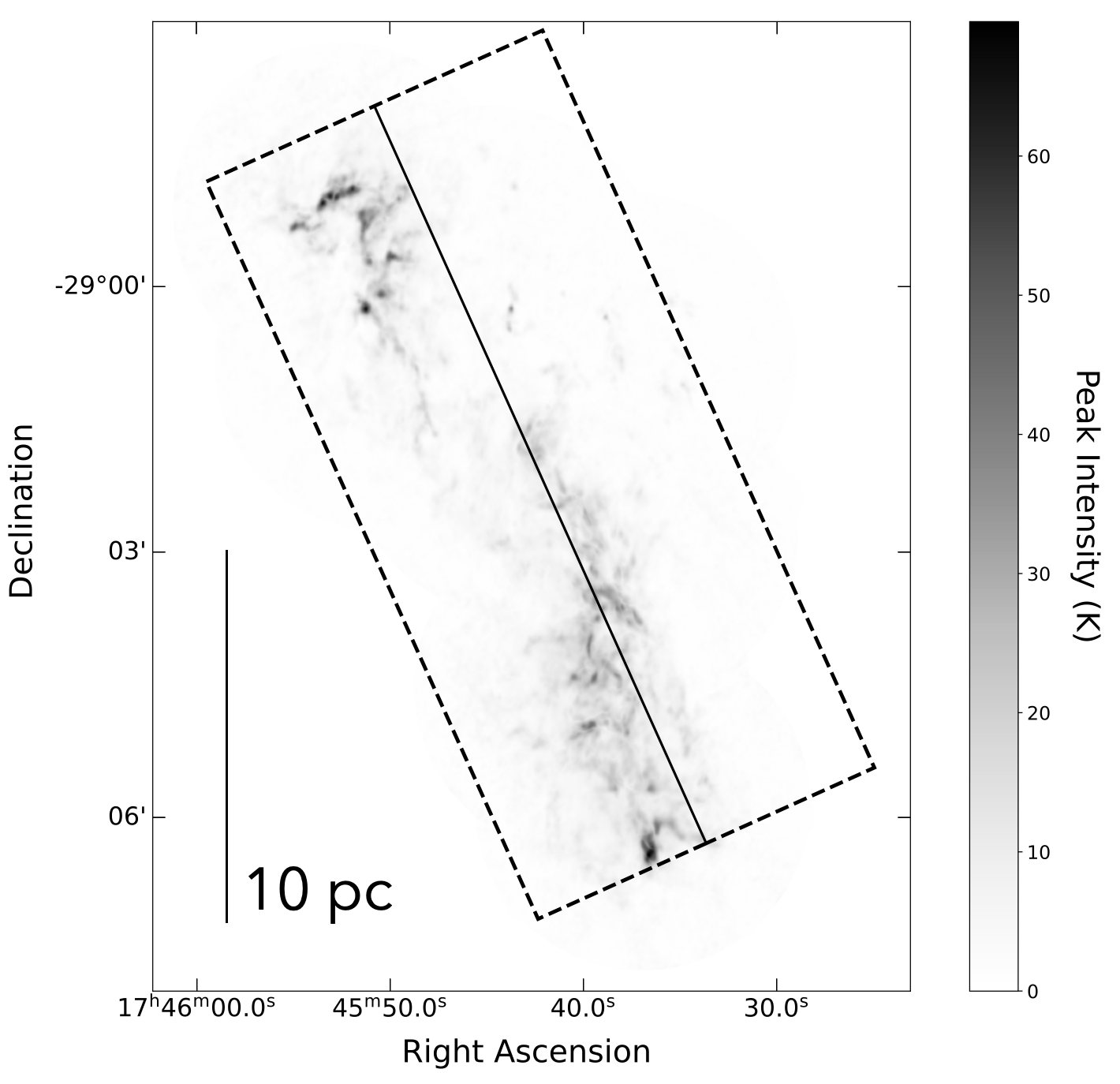}
\includegraphics[width=0.64\textwidth]{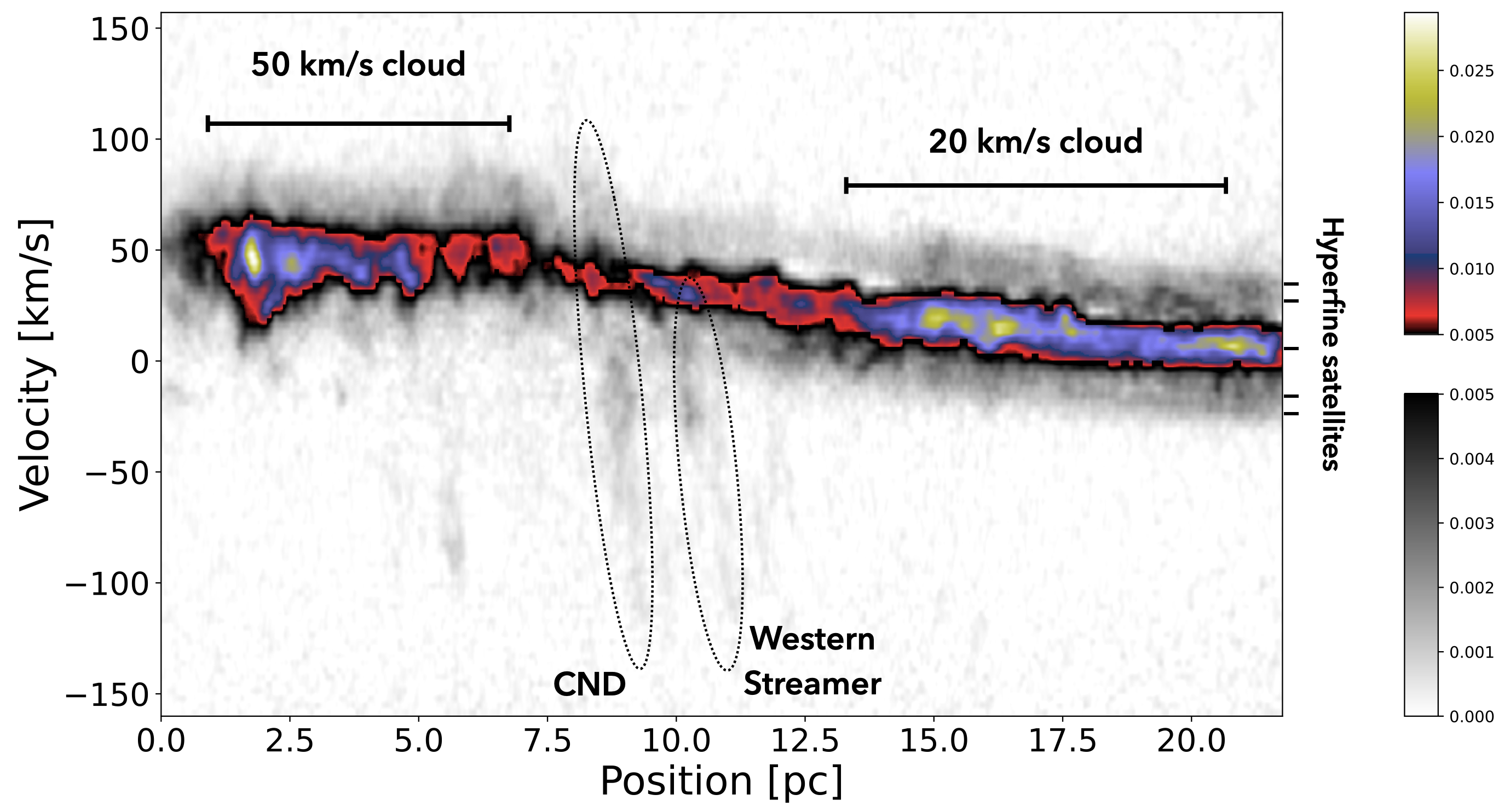}
\caption{Left: A peak intensity map of \am (3,3) overlaid with a slit of width and length used for extracting position-velocity data. Right: A position-velocity diagram of \am (3,3) observed toward the central 10 pc. The locations of prominent structure, including the 50 km/s cloud, 20  km/s cloud, CND, and `Western Streamer' are indicated. Line widths from this diagram should not be assumed to be be representative of intrinsic gas properties, due to the hyperfine structure of the (3,3) line, which is indicated (satellite lines are at $\pm$20.7 and $\pm$28.2 \kms.)} 
\label{fig:Fig-PV}
\end{figure*}

In Figure \ref{fig:Fig-velocity} we show our integrated intensity map of \am (3,3). Our seven-pointing mosaic of \am (3,3) shown in Figures \ref{fig:Fig-velocity} and \ref{fig:Fig-PV} covers a 4.5$'$ by 9$'$ (11$\times$22pc) field of view, more than doubling the survey region of \cite{McGary01} and \cite{Herrnstein05}. We include the entirety of the 50 and 20 km/s clouds, as well as the CND and the `Western Streamer'. 
While we pick out many of the same structures identified by \cite{Herrnstein05}, the enhanced resolution \citep[$2.9''\times3.7''$, compared to the Gaussian tapered beam size of $16''\times14''$ in ][]{McGary01} and sensitivity \citep[an RMS of 1.3 mJy/beam per 3.17 km/s channel, compared to an RMS of 330 mJy/beam per km/s in ][]{McGary01} of our data combined with the larger field of view provides crucial new insight into the nature of these structures and the overall distribution of gas in this region. 


The right panel of Figure \ref{fig:Fig-velocity} shows the velocity that corresponds to the brightest pixel in the cube at each position. 
As the region around the CND includes a superposition of several velocity components, this should not be assumed to be a complete portrayal of the velocity structure of the cube, since there may be multiple velocity components along the same line of sight in some positions. Still, it clearly shows the contiguous velocity structure of the 50 and 20 km/s clouds, which have a velocity gradient of $\sim$ 2.5 km s$^{-1}$ pc$^{-1}$ along the Galactic plane from 0 km/s to 60 km/s. It also shows several locations where a component of gas at a different velocity is superposed over the otherwise continuous velocity gradient of the clouds. The CND and `Western Streamer' clearly stand out as regions with larger velocity gradients, starting at -100 km/s and increasing in velocity in this image from south to north, consistent with a rotation gradient along the Galactic plane. 

In the right panel of Figure \ref{fig:Fig-velocity} we also overlay the 1720 MHz OH masers from \cite{Sjouwerman08}, illustrating how both their positions and velocities compare to the underlying gas traced with \am. OH masers associated with the CND are seen at high positive (+132 km s$^{-1}$) and negative (-104 to -141 km s$^{-1}$) velocities near the upper and lower tangent points of this ring. Other masers are seen at the periphery of the Sgr A East SNR shell. We see \am at similar velocities along the east and southeast portions of the shell, but we do not see \am at the velocities of the masers along the northwest part of the shell. 

From our \am (3,3) cube we also construct a high-resolution position-velocity (PV) diagram of the inner projected 20 pc. The slice position and width are both shown in Figure \ref{fig:Fig-PV}. The slit runs parallel to the Galactic plane. 

\begin{figure*}[tbh]
\includegraphics[width=1.0\textwidth]{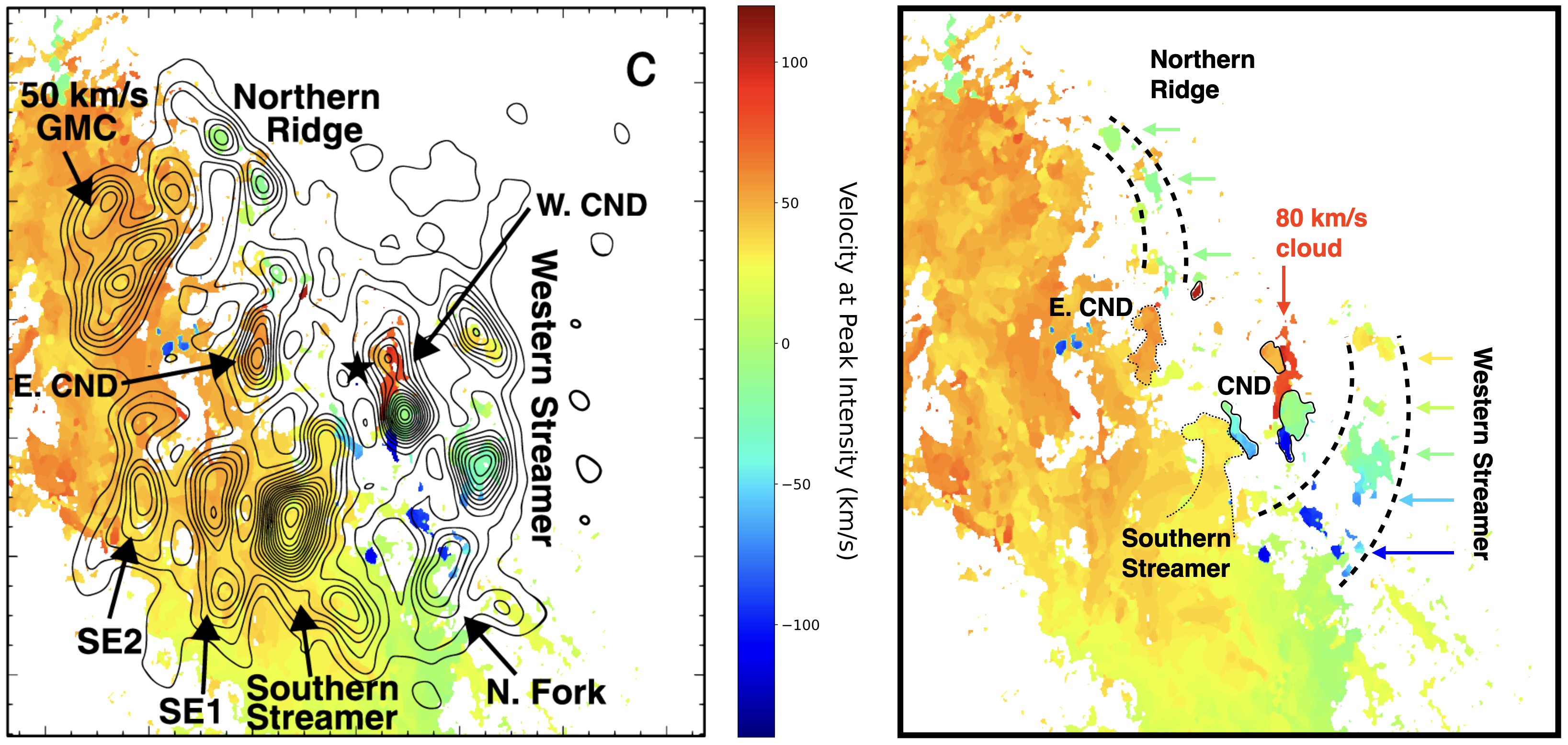}
\caption{Left: Contours of \am (3,3) emission from Figure 1 of \cite{Herrnstein05}, overlaid on the new \am (3,3) velocity data shown in Figure \ref{fig:Fig-velocity}. Right: Features from \cite{Herrnstein05} identified on our new \am (3,3) velocity data. Gas belonging to the CND is traced with solid black contours. The 80 km/s cloud, which overlaps with the CND, is not outlined. The locations of the `Western Streamer' and `Northern Ridge' are delineated with dotted black lines. As they spatially overlap with other gas, we also use colored arrows to show the velocity of the indicated feature at each position. Features that appear to be extensions of the 50 and 20 km/s clouds (the `E. CND' and `Southern Streamer', respectively) are outlined with dashed contours.  }
\label{fig:Fig-velocity-structures}
\end{figure*}

Our PV diagram of the gas observed toward the central 20 pc shows the very diverse velocity structures of previously-identified features. The CND and `Western Streamer' appear as nearly vertical structures in PV space, with a large velocity extent over a small range in position. The 50 and 20 km/s clouds in contrast have a  shallower slope. Overall, we see a smooth and continuous velocity structure connecting the 20 and 50 km/s clouds, with an average velocity gradient of 2.5 km s$^{-1}$ pc$^{-1}$, consistent with measurements by \cite{Henshaw16}, though the 50 km/s cloud has a nearly flat velocity gradient and the connecting `Molecular Ridge' has a maximum velocity gradient of 5 km s$^{-1}$ pc$^{-1}$. In contrast, the CND has a velocity gradient of 150 km s$^{-1}$ pc$^{-1}$. We note that while the CND might appear to intersect with these clouds in PV space, this is largely a result of averaging over a large region perpendicular to the slit in Galactic latitude.  We also see in the PV diagrams an additional negative velocity feature at the position of $\sim$ 6 pc along the slice. This feature can also be seen in Figures \ref{fig:Fig-velocity} and \ref{fig:Fig-velocity-structures} as a dark blue (high negative velocity) structure superposed on the western edge of the 50 km/s cloud. The velocities of this gas are 'forbidden' for circular orbits, implying that this gas must be on a highly non-circular orbit. We do not propose an explanation for these kinematics or a location for this gas, but we note that previous observations of this feature have suggested it could be associated with Sgr A East, as it is projected to lie within the SNR shell \citep{Genzel90,Zylka99}.

In Figure \ref{fig:Fig-velocity-structures} we show the features introduced in the \citetalias{Herrnstein05} model overlaid on our peak velocity map from Figure \ref{fig:Fig-velocity}. We also note the presence of an additional feature known as the 80 km/s cloud \citep[or 70 km/s cloud;][]{Jackson93}, which spatially coincides with the western edge of the CND, yet does not show the large rotational velocity gradient typical of gas in the CND \citep{Karlsson15} or any sign of the ionization seen in the western edge of the CND \citep{Christopher05} suggesting that it may be separated from the CND along the line of sight. Two of the features, in the \citetalias{Herrnstein05} model, the `E. CND' and the `Southern Streamer', appear to be consistent with the overall velocity field of the 50 and 20 km/s clouds. \cite{Coil99} initially defined the `Southern Streamer' as a 2 pc by 10 pc feature with a velocity gradient that they interpreted as infalling motion toward the CND. Figure \ref{fig:Fig-velocity-structures} only shows the northern tip of this structure, the southern extension of which is a main component of the `Molecular Ridge' and the 20 km/s cloud. Indeed, the velocity gradient described by \cite{Coil99} of 2-3 km s$^{-1}$ pc$^{-1}$ appears to just correspond to the overall orbital velocity gradient of the 20 km/s cloud (2.5 km s$^{-1}$ pc$^{-1}$). Similarly, although the `E. CND' or `Northeast arm' \citep{Christopher05} is often treated as part of the CND because of its apparent proximity \citep[e.g., ][]{Stel23}, our velocity map shows that it is more likely to be an outer part of the 50 km/s cloud. In the \citetalias{Herrnstein05} model and its later updates, the `Northern Ridge', `Western Streamer',  and `Southern Streamer' are all suggested to be features that connect the 50 and 20 km/s cloud to the CND. Contrary to these models, we do not see any sign that the `Northern Ridge' connects either to the 50 km/s cloud or the CND. The `Northern Ridge' can be seen to overlap spatially with the 50 km/s cloud in Figure \ref{fig:Fig-velocity}, but consists of gas at distinctly separate velocities (-15 km/s to -5 km/s, compared to 40 to 60 km/s in the nearby 50 km/s cloud and `E. CND'). It may also continue beyond the extent of the 50 km/s cloud: we see additional gas at these velocities at the nothernmost extent of the CND, visible in the top left of the second panel of Figure \ref{fig:Fig-velocity-structures}. While the `Northern Ridge' also comes close in projection to the northern part of the CND, gas there has velocities of +100 km/s.  We also do not see any evidence that the `Western Streamer' connects to the 20 km/s cloud \citep[as is implied by the updated model of ][]{Lee08}. At the location where they spatially overlap, the streamer velocities are -110 to -90 \kms, while the cloud velocity is 0 to 20 \kms. Neither do we see any sign of a connection of the `Southern Streamer' with the CND, which is also a feature of the \cite{Lee08} model.  While the northern extent of the `Southern Streamer' abuts the southeast part of the CND, the velocities are again discontinuous: the `Southern Streamer' has velocities of 20 - 30 \kms, while the nearby CND has velocities of -70 to -30 \kms.

From this we infer the following: 
\begin{itemize}
    \item The `E. CND' is not an extension of the CND, but rather appears to be consistent with the velocity field of the 50 km/s cloud. 
    \item The `Northern Ridge' is not associated with the CND or 50 km/s cloud. {Its velocity (near 0 km s$^{-1}$) is consistent with relatively local foreground gas, but its detection in the relatively highly-excited \am (3,3) line makes this association somewhat unlikely. This velocity could also be consistent with the 20 km/s cloud, though it is relatively far displaced ($>$5 pc) from other gas in the cloud at this velocity, so we suggest this is unlikely as well. Most likely it is kinematically-disturbed gas in the inner kiloparsec, perhaps even lying between the the `x2' orbits and the CND, but we cannot further constrain its position.} The 50 km/s cloud is not feeding the CND through this feature.
    \item The `Western Streamer' is not connected to the 20  km/s cloud. Its kinematics appear consistent with the rotation signature of the CND, and we conclude that this is consistent with being associated with the outer part of the CND, potentially located as close to Sgr A* as it appears in projection ($r\sim$ 3-5 pc).
    \item The `Southern Streamer' is a part of the 20 km/s cloud that appears close in projection to the CND, but is not connected to the CND. The 20 km/s cloud is not feeding the CND through this feature.

\end{itemize}

\subsection{Chemistry of the CND and surrounding gas}

\begin{figure*}[tbh]
\includegraphics[width=1.0\textwidth]{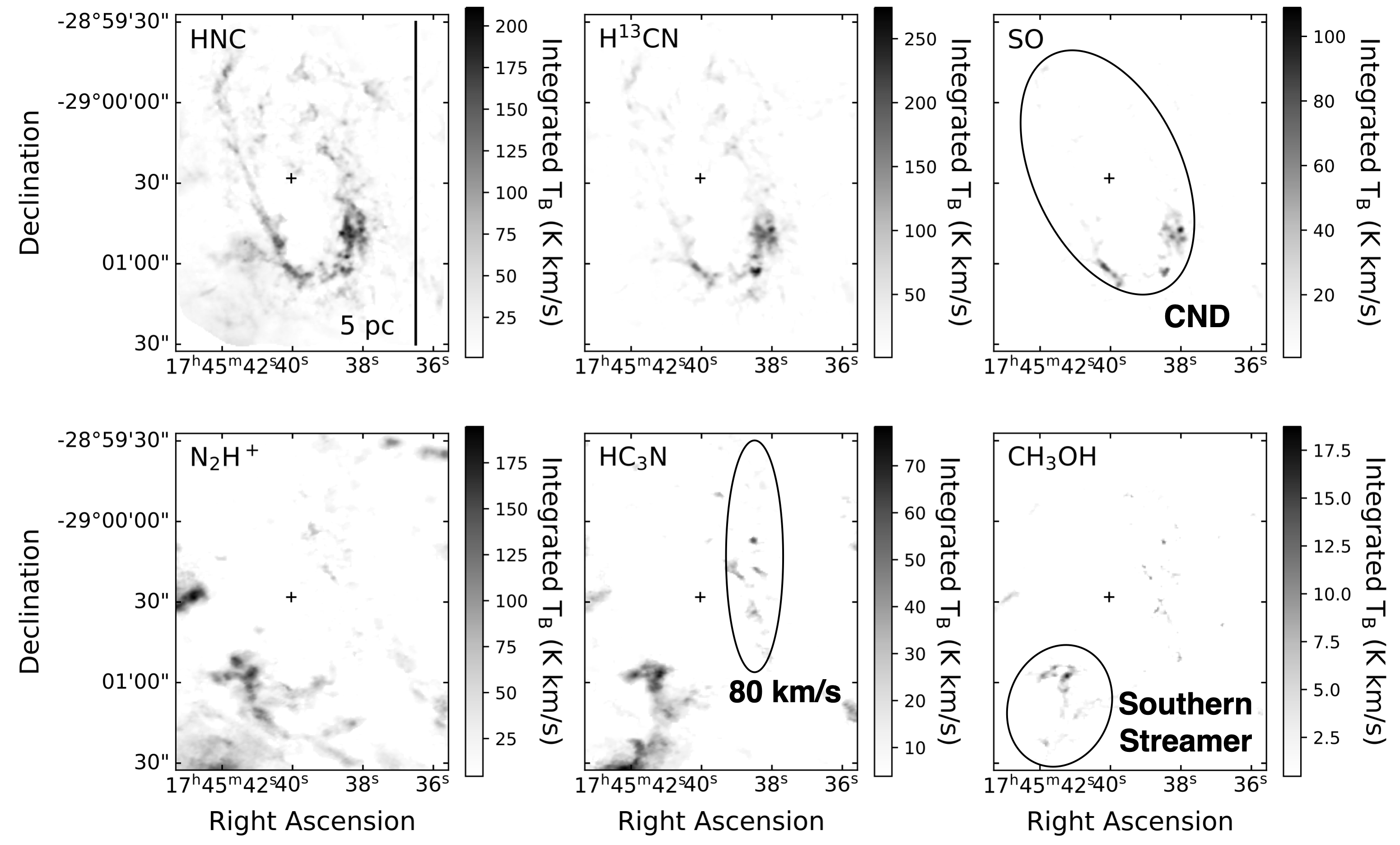}
\caption{Maps of integrated intensity toward the CND for lines in Table \ref{tab:ALMA}. Key features are labeled, and the location of Sgr A* is marked with a black cross. }
\label{fig:Fig-cnd-int-intensity}
\end{figure*} 

\begin{figure*}[tbh]
\includegraphics[width=1.0\textwidth]{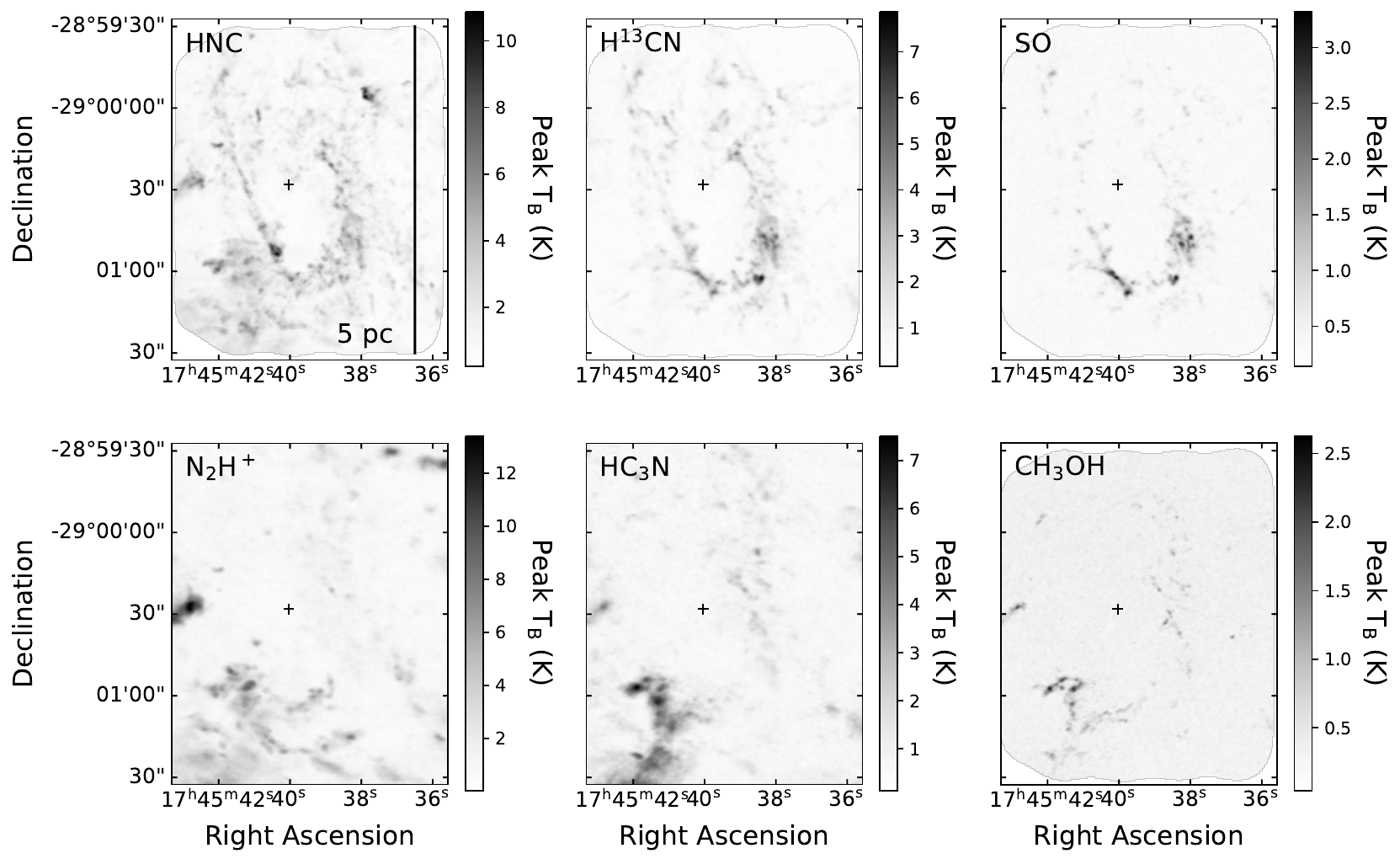}
\caption{Maps of peak intensity toward the CND for lines in Table \ref{tab:ALMA}. The location of Sgr A* is marked with a black cross. }
\label{fig:Fig-cnd-peak-intensity}
\end{figure*}

\begin{figure*}[tbh]
\includegraphics[width=1.0\textwidth]{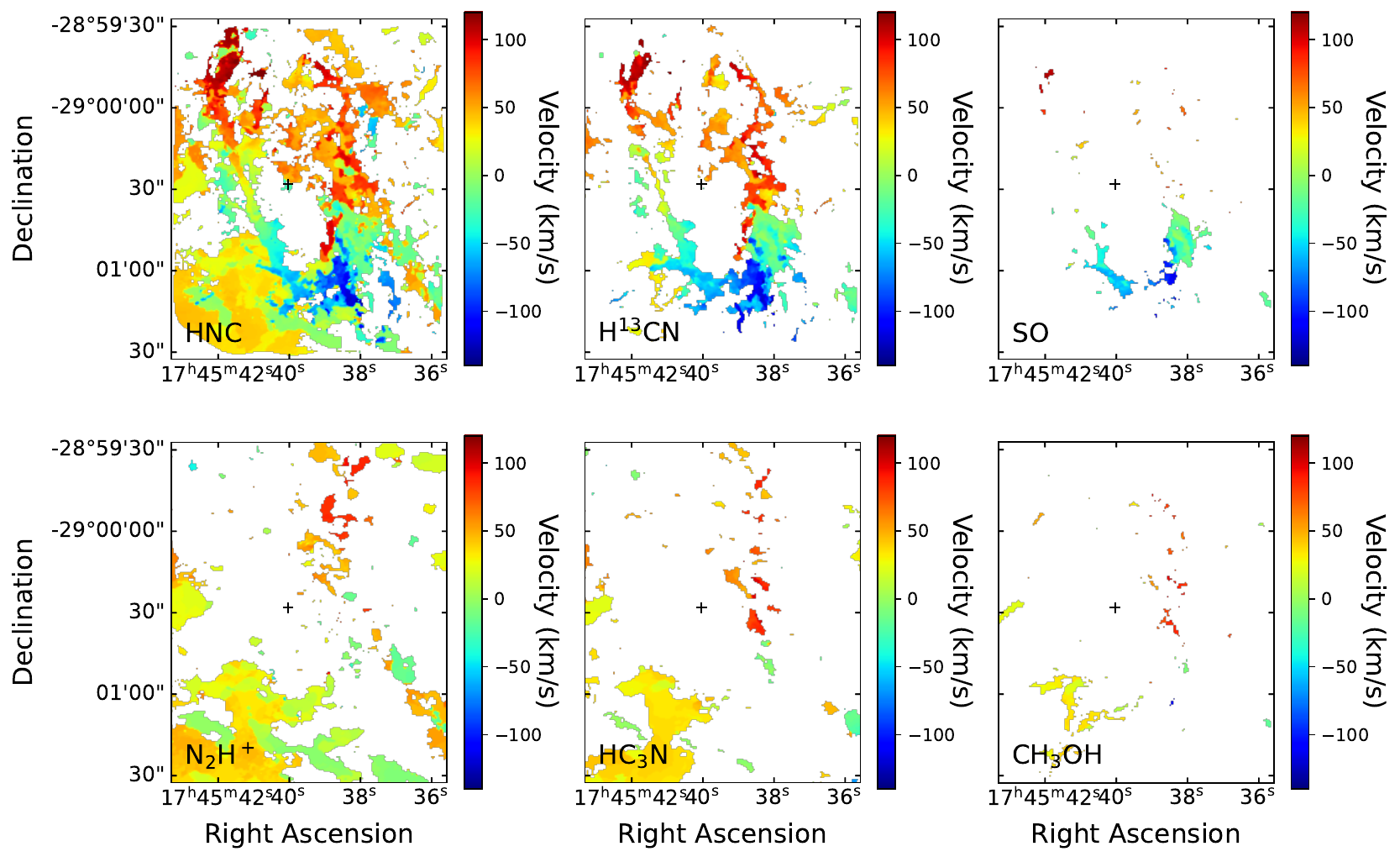}
\caption{Maps of the velocity value corresponding to the peak intensity for each pixel. The velocity scaling is the same as in Figures \ref{fig:Fig-velocity} and \ref{fig:Fig-velocity-structures}. The location of Sgr A* is marked with a black cross. Values in the cube below 5 times the noise given in Table \ref{tab:ALMA} were masked, except for CH$_3$OH which was masked at the 3 sigma level.} 
\label{fig:Fig-cnd-peak-vel}
\end{figure*} 

In addition to \am, we present maps of 6 representative lines (listed in Table \ref{tab:ALMA}) from our ALMA observations of the CND, illustrating the chemical differentiation between gas in the inner CND and other gas in the same field of view. In Figure \ref{fig:Fig-cnd-int-intensity} we show integrated intensities from all of the lines, in Figure \ref{fig:Fig-cnd-peak-intensity} we show the peak intensities (the brightest pixel in the cube at each position) from all of the lines, and in Figure \ref{fig:Fig-cnd-peak-vel} we show the velocity that corresponds to each of these peak intensities, on the same scale as Figure \ref{fig:Fig-velocity}. We see striking variations in the distributions of molecular emission in and around the CND. While emission from HNC (3-2) is broadly present in all of the gas components, the strongest emission from the inner CND is seen in H$^{13}$CN (3-2)  and SO ($6_5-5_5$). In contrast, the `Southern Streamer' and other gas at the velocity of the 50 and 20 km/s clouds is most pronounced in CH$_{3}$OH, HC$_3$N, and N$_2$H$^+$. An additional feature not present in the \citepalias{Herrnstein05} model, `the 80 km/s cloud' \citep{Jackson93,Karlsson15}, can also be seen in Figures \ref{fig:Fig-cnd-int-intensity} and \ref{fig:Fig-cnd-peak-vel}, where it is visible in all tracers except SO. 

We note that the variation we see between the observed molecules is not just an effect of excitation, though this likely plays some role \citep[gas in the CND is substantially warmer and denser than gas in the CMZ;][]{Mills13,Mills17b,Mills18a}, as the observed transitions of CH$_3$OH and SO have similar upper level energies $E_{upper}$. Instead, this chemical differentiation is consistent with prior work by \citet{Martin12} whose analysis suggested that molecules more susceptible to photodissociation (e.g., H$_2$CO and HC$_3$N) are not abundant in the molecular gas of the inner CND, likely due to proximity to the radiation field from the young component of the nuclear cluster. Our results reinforce this finding: not only do we also see that HC$_3$N is faint in the CND compared to surrounding gas, we also see this for CH$_3$OH, another molecule known to have lower abundances in photon-dominated regions \citep[PDRs, e.g., M82;][]{Martin06}. Compared to H$^{13}$CN, the HNC emission is relatively weaker in the CND compared to the surrounding clouds. This could be consistent with HNC being photodissociated faster than HCN at comparable temperatures \citep{Aguado17}. N$_2$H$^+$ also appears largely absent in the CND, which could either be as a consequence of dissociative recombination with free electrons, or increased reactions with CO in this dense gas to form HCO$^+$ and N$_2$, as is suggested to be the case for the dense gas in M82 \citep{Mauersberger91}. The latter scenario is supported by the existence of an additional warm dust component in the CND, which is observed to have two dust temperature components of 23 K and 45 K \citep{Etxaluze11}. This can be compared to the CMZ, for which the median dust temperatures are 21$\pm$4 K \citep{Battersby25a}. The higher dust temperatures in the CND would then lead to less freeze-out of CO, increasing the gas-phase CO available for reactions with N$_2$H$^+$. Alternatively, \cite{SantaMaria21} argue that a high ionization rate environment, particularly due to cosmic rays (as applied to the CMZ cloud Sgr B2) will make N$_2$H$^+$ generally abundant even in warm gas as long as it is dense (n$>10^4$ cm$^{-3}$) and highly-irradiated (cosmic ray ionization rates in excess of $10^{-15}$ s$^{-1}$). However, this does not explain the relative dearth of N$_2$H$^+$ toward the CND, where gas densities are also high \citep[n$>10^4$ cm$^{-3}$][]{RequenaTorres12,Mills13} and the cosmic ray ionization rate is inferred to be large \citep[$>10^{-15}$ s$^{-1}$][]{Harada15}.    

Not only does the `Southern Streamer' appear chemically  distinct from CND gas, the velocities of gas in this feature as measured in the ALMA tracers is also not consistent with apparently nearby or overlapping gas in the CND (though the velocities are consistent with the 20 km/s cloud, as described in Section \ref{sec:Ammonia}). This further supports locating this feature farther than from Sgr A* than it appears to be in projection. The same is true of the 80 km/s cloud \citep{Jackson93,Karlsson15}, which clearly lies at a distinct velocity from CND gas at the same position, which has velocities from -20 to 50 km/s. In this case of the 80 km/s cloud, the lack of any ionized gas observed at this velocity \citep{Zhao09}, coupled with the abundance of molecules like CH$_3$OH and HC$_3$N that are easily photodissociated, indicates the 80 km/s cloud is clearly subject to a less intense radiation field than the CND, and likely some distance in front or behind this structure. A more thorough study of the ALMA data, examining differences in chemistry and kinematics between the CND and surrounding gas, will be presented in a companion paper (Mai et al. in prep.).

\subsection{CO absorption toward Sgr A*}
\begin{figure*}[tbh]
\includegraphics[width=0.31\textwidth]{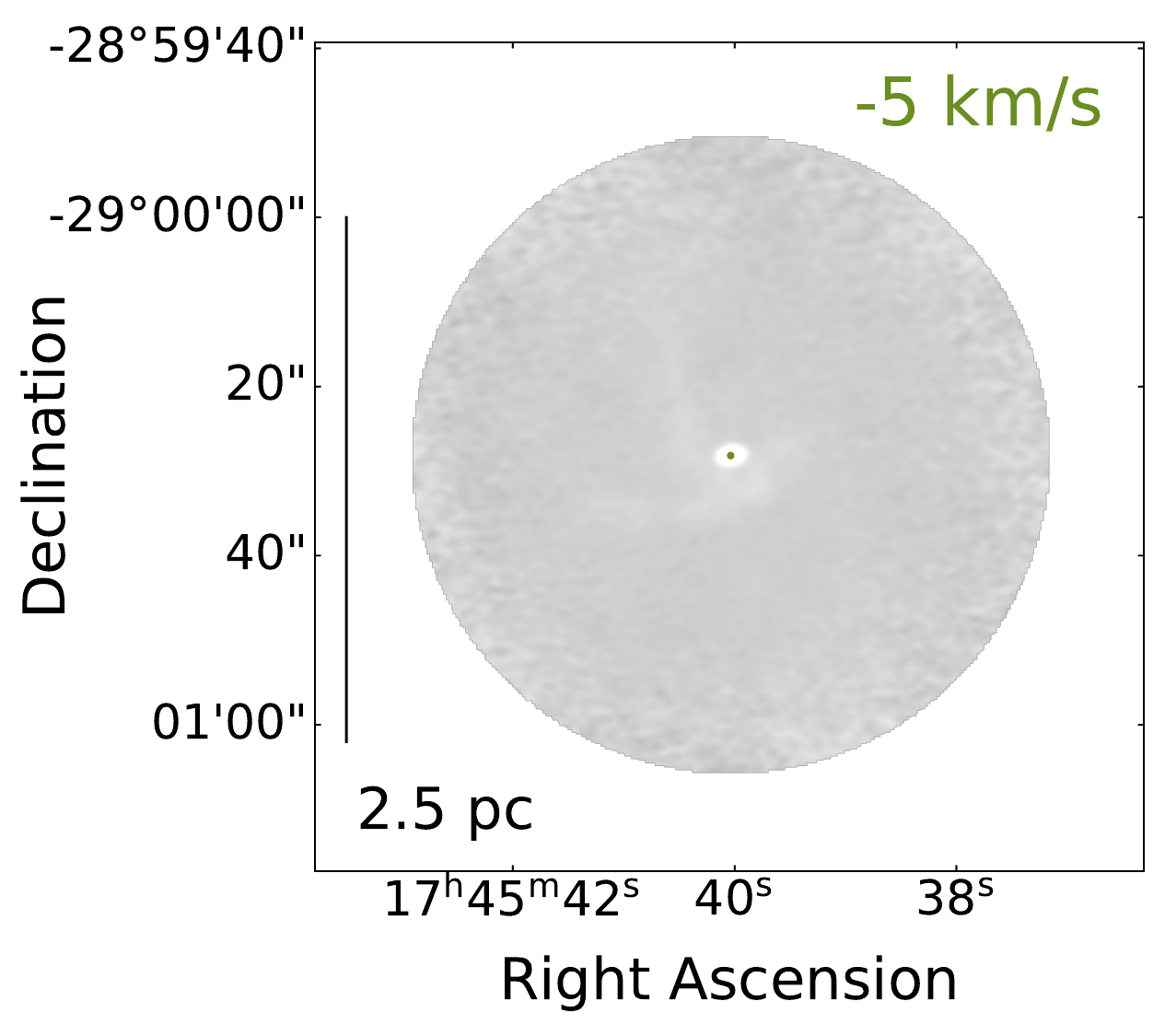}\includegraphics[width=0.31\textwidth]{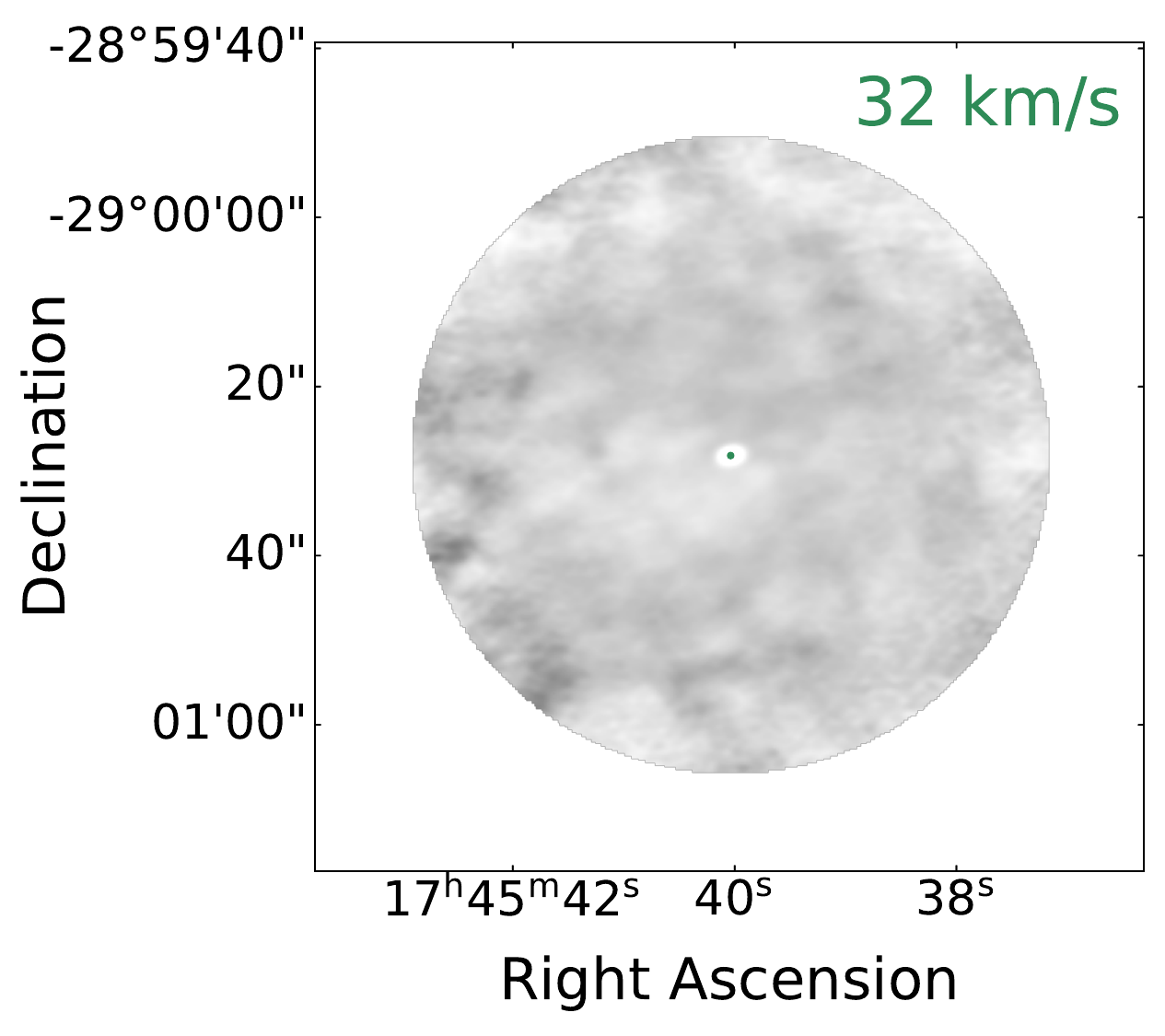}\includegraphics[width=0.37\textwidth]{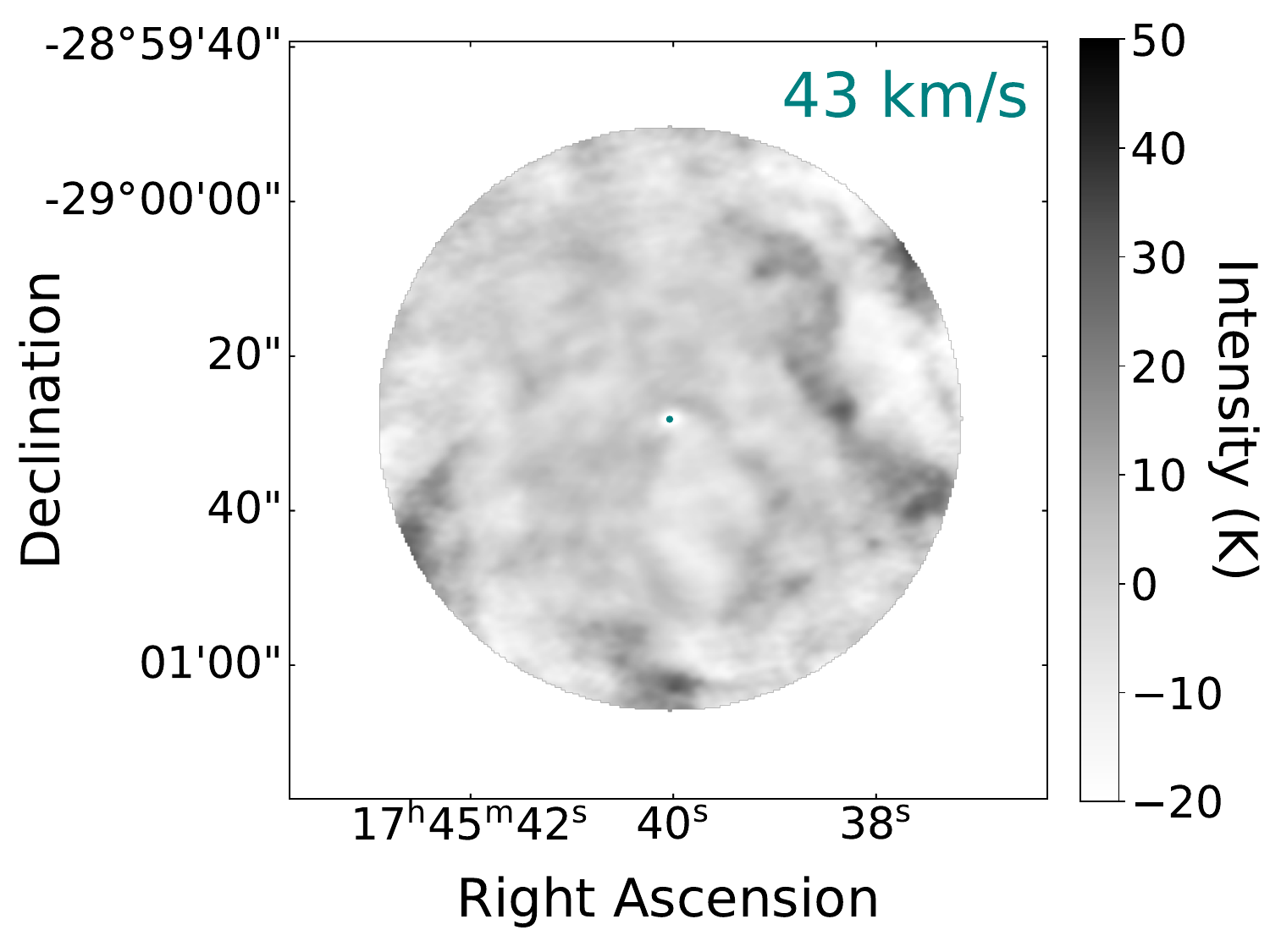}\\
\includegraphics[width=0.31\textwidth]{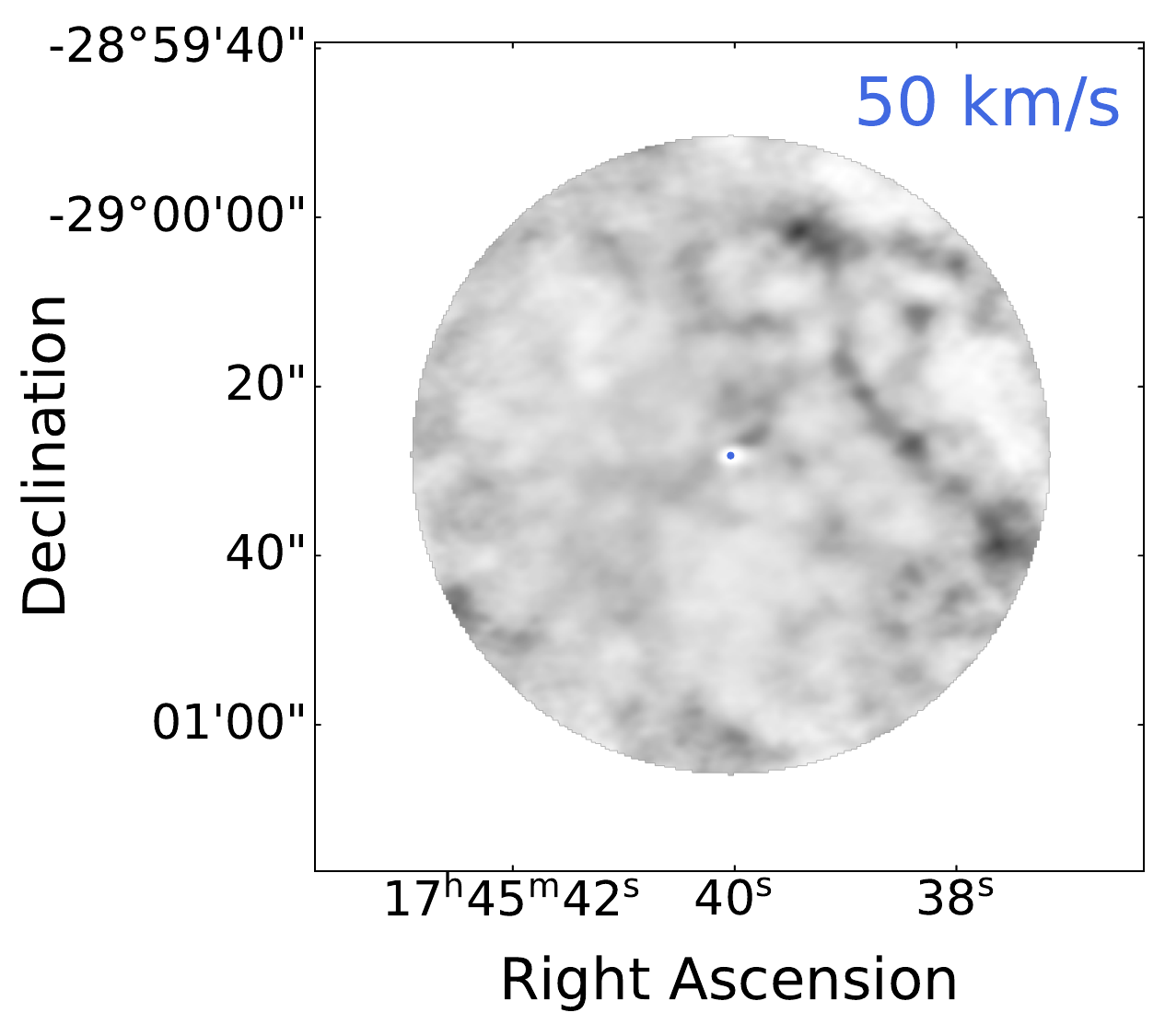}\includegraphics[width=0.31\textwidth]{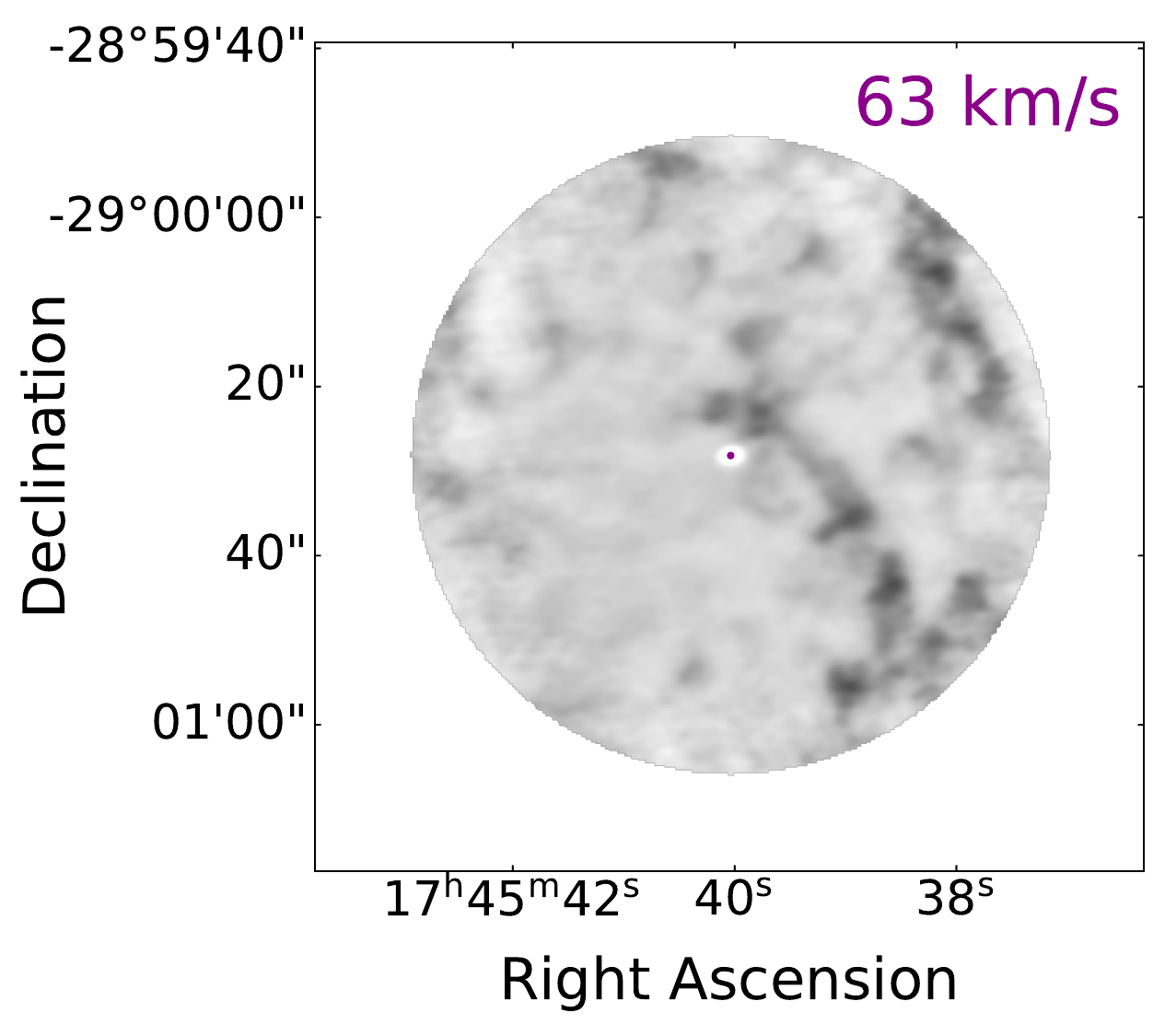}\includegraphics[width=0.37\textwidth]{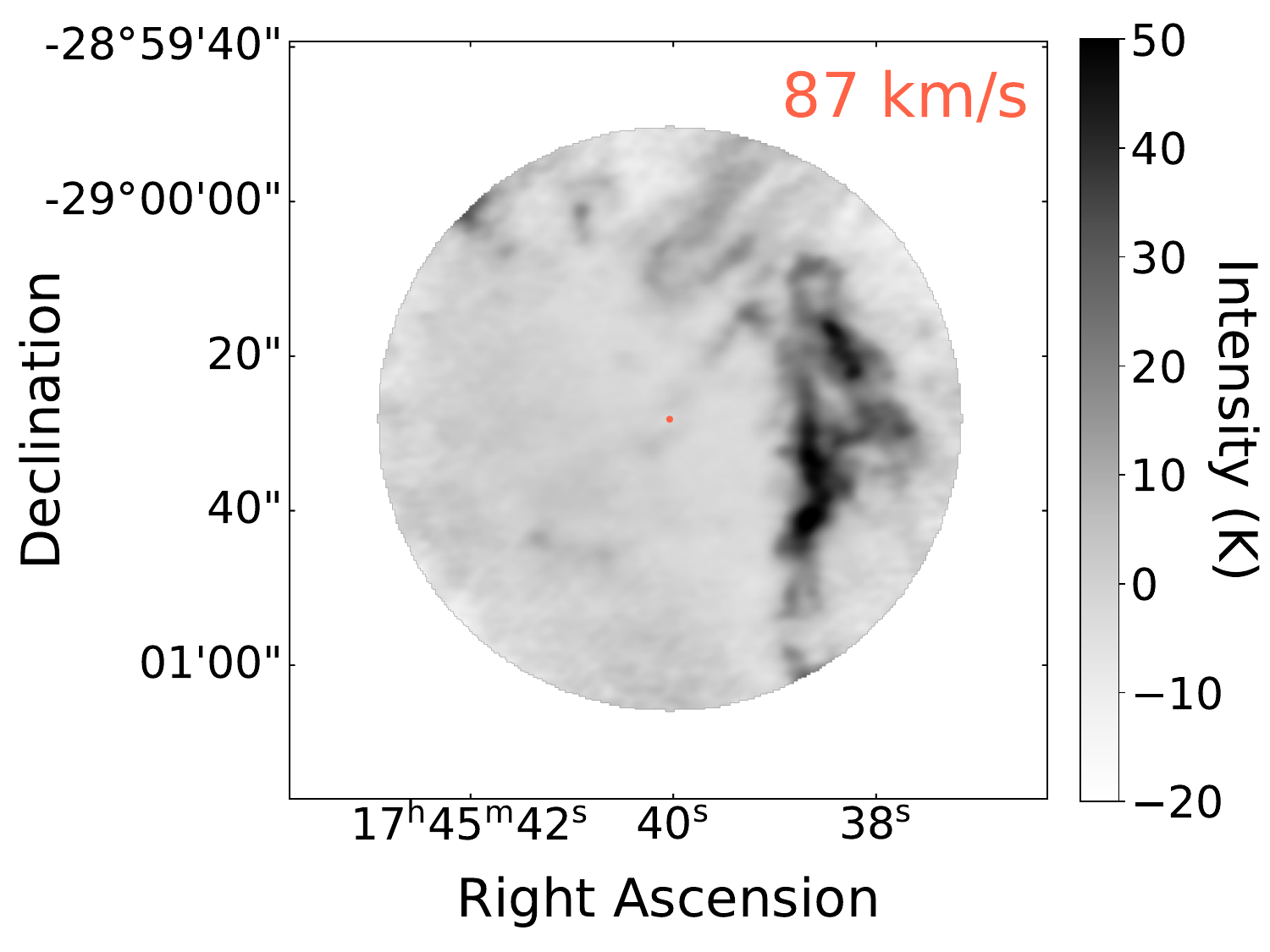}
\\
\includegraphics[width=1.0\textwidth]{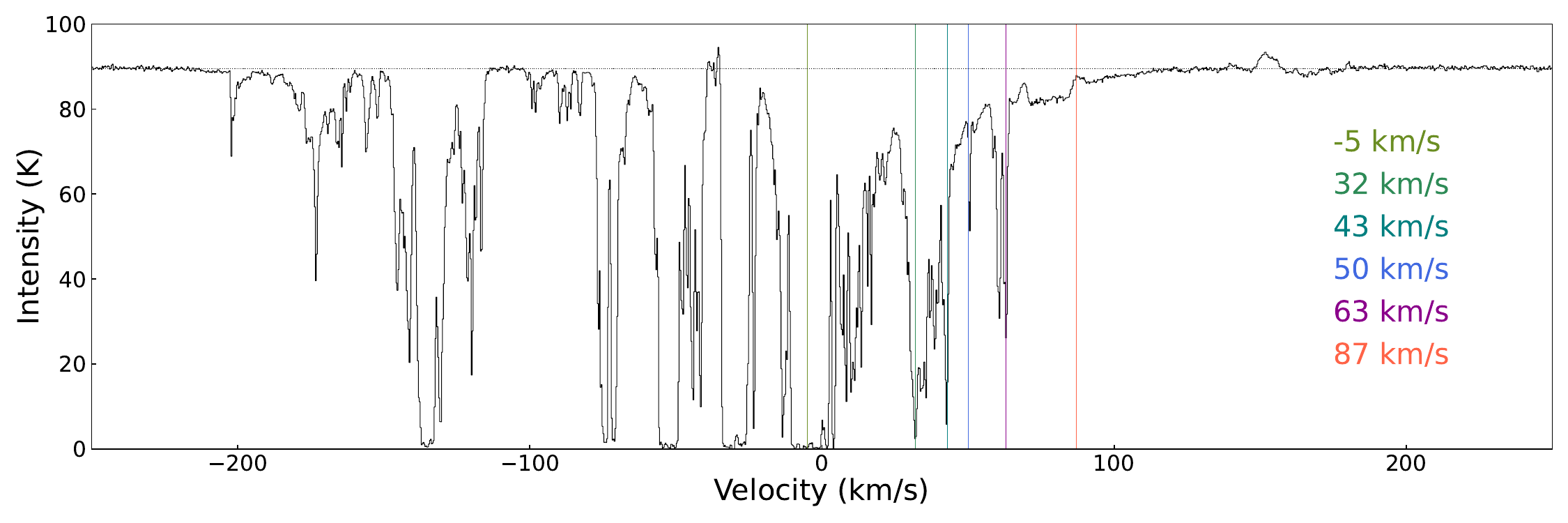}
\caption{Top panels: Channel maps of \co ($J=1-0$) for selected velocities indicated in the spectrum below. The position of Sgr A* is marked with a circle in each panel. Bottom Panel: Absorption spectrum of \co ($J=1-0$) extracted from a single pixel at the position of Sgr A*. Six velocities of interest are marked with vertical lines.}
\label{fig:Fig-absorption}
\end{figure*} 

We measure an absorption spectrum of \co $J=1-0$ toward the unresolved bright (90 K / 3.3 Jy at 115 GHz) point source Sgr A*. As \co is the most abundant species in the ISM after H$_2$ and absorption in the $1-0$ transition probes CO in the highly-populated ground rotational state, we expect that this spectrum contains absorption from gas not just in the CMZ but along the full line of sight from the sun to Sgr A*. Gas behind Sgr A* (in the back half of the CMZ) will not be present in absorption in this spectrum. 

The resulting \co absorption spectrum toward Sgr A* (Figure \ref{fig:Fig-absorption}) shows dozens of velocity components ranging from -200 km/s to 63 km/s that lie in front of Sgr A*. The strongest absorption features (which are saturated in our spectrum) all occur at negative velocities. We also see superposed on these narrower features a broader envelope of absorption from -100 to 100 km/s. While gas within this velocity range could originate either within the CND or in the front half of the CMZ, \cite{Dwarakanath04} interpret a similar broad feature with a FWHM of 113 km/s seen in HI as due to gas local to the CND, on the basis of a position-velocity analysis of this feature. 

Here, we focus specifically on the velocity range of the 50 and 20 km/s cloud. We identify 4 narrow (FWHM $\sim 1-2$ km/s) absorption features superposed on the broader absorption trough: 
\begin{enumerate}

\item 32 km/s. The velocity of the `Southern Streamer'
\item 43 km/s. A velocity consistent with gas in the `Molecular Ridge' between the 20 and 50 km/s clouds
\item 50 km/s. A narrow and isolated component in the velocity range of the 50 km/s cloud
\item 63 km/s. Part of a double trough feature in the velocity range of the 50 km/s cloud, but also at the velocity of the OH streamer observed by \citep{Karlsson15}

\end{enumerate}

While Sgr A* lies relatively far from the cloud centers, we do see spatially-extended gas features in N$_2$H$^+$ (Figures \ref{fig:Fig-cnd-int-intensity} and \ref{fig:Fig-cnd-peak-vel}) that overlap the CND at similar velocities. This indicates that it is plausible for extended gas in the periphery of these clouds to be overlapping with Sgr A*. \cite{Karlsson15} also identify an OH absorption feature between 40 and 67 km/s that may be related to the features at 50 and 63 km/s. While \cite{Karlsson15} relate it to the CND based primarily on its spatial location, it is not obviously part of this feature, and appears to connect in their channel maps to extended absorption from the 50 km/s cloud. In general, we note that gas clumps in the CND are characterized by relatively large turbulent line widths (FWHM $\sim 11-55$ km/s) at comparable spatial resolutions of a few arcseconds \citep{Christopher05,MonteroCastano09}, while these features have line widths of 1-2 km/s. Furthermore, we don't expect much overlap of CND gas with Sgr A* given that the CND's central cavity is largely evacuated of molecular gas \citep{Christopher05}, the inclination of the CND is more face-on than the rest of the CMZ \citep[67-80$\degr$;][]{Martin12,Lau13} and the disk has a relatively small thickness \citep[0.34 pc;][]{Lau13}. Ultimately, based on their linewidth and velocity, we then identify these four absorption features as being associated with the 50 and 20 km/s clouds, which must then lie in front of Sgr A*. 
This is consistent with the overall choice of geometry in most recent kinematic models \citep[e.g.,][and Lipman et al. (in prep).]{Kruijssen15,Sofue25}, as well as the proper motion observations of \citep{NoguerasLara26} which place the clouds $\sim$ {35} - 70 pc in front. 

We also compare these features to two other features in the spectra: absorption at -5 km/s (an example of a saturated absorption profile) and a feature at 87 km/s (a local maximum in the spectrum superposed with the overall broad absorption envelope) that is likely associated with the 80 km/s cloud. For the saturated absorption at -5 km/s, the channel map shows absorption against the minispiral as well as Sgr A*. The image otherwise appears largely featureless, indicating emission from the absorbing gas likely arises on large spatial scales such that it is resolved out by the interferometer. This suggests that this feature corresponds to more local gas. In contrast, we see clear emission structure in the channel maps at the velocity of the other highlighted absorption features. For the 87 km/s feature, we see narrow superposed emission (as opposed to superposed absorption) at this velocity in the small aperture toward Sgr A*. While this is suggestive of the 80 km/s cloud feature lying behind Sgr A*, there remains a possibility that the emission comes from a clumpy feature that is both smaller than the $2.4'' \times 1.6''$ beam, and not spatially overlapping with Sgr A*.

\section{Discussion}
\label{discussion}

We have analyzed a new map of \am 3,3 emission in the central $11 \times 22$ pc as well as  new ALMA data toward smaller fields of view (a 3$\times$4 pc mosaic and a 2 pc radius pointing). We do not observe any connections in position-velocity space between the CND and either the 50 km/s or 20 km/s clouds. We see a clear chemical difference for gas associated with the inner CND and gas at velocities of the 50 and 20 km/s clouds, including the tip of the `Southern Streamer'. We also see \co absorption at the velocities of the 50 and 20 km/s clouds, which strongly suggests that both lie in front of Sgr A*. Overall, our analysis shows that there is no evidence for a connection between the CND and the apparently nearby 20 and 50 km/s clouds. The CND and `Western Streamer' (which largely lies outside of our ALMA maps), appear to be the only molecular gas located within a radius of 3 pc from the supermassive black hole. While our absorption measurements alone cannot determine how far in front the 20 and 50 km/s clouds lie, kinematic models (Table \ref{tab:models}) as well as recent independent geometric constraints \citep[][and Lipman et al. submitted]{NoguerasLara26} place these clouds at least {15} - 20 pc in front, inconsistent with the \citetalias{Herrnstein05} model. 

The primary issue remains: the inferred line-of-sight distances between the \citetalias{Herrnstein05} model and the kinematic models are not consistent. Point by point, the existing observational requirements are that Sgr A East is behind Sgr A* and the Sgr A West minispiral {(Section \ref{sec:ii})}, that Sgr A East is directly interacting with the 50 km/s cloud and possibly Sgr A West {(Sections \ref{sec:iii}, and \ref{sec:iv})}, and that the 50 km/s cloud (and 20 km/s cloud) both lie in front of Sgr A* and the Sgr A West minispiral. Simply requiring the 50 km/s cloud to be in front stresses the \citetalias{Herrnstein05} model but does not fully break it: the 50 km/s cloud can lie slightly in front of Sgr A East while the  bulk of the SNR shell is behind (or enveloping) Sgr A West. However, if the 50 km/s cloud lies more than 5 pc in front of Sgr A*, {the suppositions described in Sections \ref{sec:ii}, \ref{sec:iii}, and \ref{sec:iv}} can no longer be satisfied simultaneously. 

We now critically revisit the original evidence and constraints that have shaped these models to search for any ways to reconcile these apparently opposing requirements.

\subsection{Revisiting the Constraints on Sgr A East’s Interaction with its Surroundings}

We first address the strength of the evidence for the interaction of Sgr A East and the 50 km/s cloud. Sgr A East is just one of a whole class of mixed-morphology SNRs believed to be interacting with {the ambient interstellar medium} \citep{Frail96,Reach96,Koo97,Rho98}. A defining characteristic of {interactions between SNRs and molecular clouds} is a type of maser that appears to be exclusively produced in an X-ray irradiated shock environment where a supernova remnant impacts a nearby cloud \citep{Frail96,YusefZadeh03}. It has been well established that other SNRs in this class show the same pattern of maser emission on their periphery, and the locations of these masers correspond to nearby molecular clouds \citep{Frail96,Frail98,YusefZadeh03}. One could attempt to argue that while Sgr A East must then be interacting with some surrounding gas, that gas does not necessarily have to be part of the 50 km/s cloud. However, this would require a coincidence of not only position but cloud morphology \citep[the cloud is is curved and apparently compressed where it touches the Sgr A East shell;][]{Serabyn92}, gas velocity (the masers are at the same velocity as the 50 km/s cloud, as can be seen in Figure \ref{fig:Fig-velocity}), and perhaps even supernova morphology (as Sgr A East appears brightest on the side that appears to be interacting with 50km/s). While we cannot fully rule out the possibility that Sgr A East and the 50 km/s cloud are unrelated, we view this as highly unlikely. 

A separate issue is whether Sgr A East is interacting with Sgr A West. The evidence for this is more circumstantial, as it depends largely on the morphological interpretation of X-ray features. The coincidence of the western edge of Sgr A East is invoked \citep{Baganoff03}, as well as the interpretation of a ridge of x-ray emission as a bow shock from the interaction of stellar winds from the nuclear cluster with supernova ejecta \citep{Maeda02,Rockefeller05,Zhang23}. No alternative explanations for these features are presented in the literature, and thus the possibility of this interaction can not be easily dismissed. However there is also a wealth of complex physics in the central parsecs: stellar winds and a young nuclear cluster old enough to have experienced supernovae \citep{Lu13}, strong magnetic fields \citep{Morris03,Ponti19}, and many independent suggestions of signatures corresponding to black hole winds or outflows \citep[e.g.,][]{Li13,Zhu19,YusefZadeh20,Cecil21}. We suggest that alternatives to interaction with Sgr A East could be further explored in order to better assess the robustness of this claimed interaction.

In summary, we find that the evidence that Sgr A East and the 50 km/s cloud are connected is strong and convincing. The evidence that Sgr A East is interacting with the central parsec is compelling, but may still leave room for an alternative interpretation.

\subsection{Revisiting Orbital Constraints}



\cite{Kruijssen15} constrain the line of sight distance of the 50 and 20 km/s clouds by fitting their orbital Stream 1 to an apparently continuous feature spanning 100 pc in Galactic longitude. They argue that allowing a pericenter approach $<$ 40 pc is not consistent with this overall orbital fitting. Reducing the pericenter approach while keeping the 50 and 20 km/s clouds connected to the larger structure of Stream 1 causes offsets in the best-fit orbits (their Figure 7) that are a poor fit to the observed velocities of all of the clouds on Stream 1. An important takeaway from the non-circular orbital fitting done by this work is that reducing the pericenter distance for the 50 and 20 km/s clouds steepens the observed velocity gradient across the clouds.  Assuming the potential adopted by \cite{Kruijssen15} is broadly correct for the inner 10 parsecs \citep[though an improved potential for the nuclear cluster has since been presented in][]{Feldmeier17}, then a pericenter distance of 10 pc \citep[][Figure 7]{Kruijssen15} requires a steeper line-of-sight velocity gradient for these two clouds that is inconsistent with the observations. Indeed, striking differences can be seen in our PV diagram when comparing these clouds with the CND. The CND and `Western Streamer' exhibit a steep velocity gradient of 130 km s$^{-1}$ pc$^{-1}$ for gas which is clearly orbiting Sgr A* at a radius of 1.4 - 3 pc (for the CND) and orbiting with a projected distance of 3 - 4.5 pc (for the `Western streamer').  In contrast, the 20 and 50 km/s clouds (with projected distances of 4.5 - 6.5 pc) exhibit a continuous, shallow slope of 2.5 -5 km s$^{-1}$ pc$^{-1}$.  Overall, the somewhat poor fit of the \cite{Kruijssen15} model to the 50 and 20 km/s clouds, which exhibit a slightly steeper velocity gradient than the overall Stream 1 orbit, does leave open the possibility that these clouds are part of a structure that is more complicated than a simple, single orbital stream.

\cite{Tress20} attempt to explain the poor fit of the \cite{Kruijssen15} model in the region of the 50 and 20 km/s clouds, based on comparison of Milky Way data with a hydrodynamic simulation. They suggest that the 50 and 20 km/s clouds are on a spur in which the stream identified by \cite{Kruijssen15} splits from an orbit with a $\sim$100 pc radius to join an inner 50 pc radius disk, and that the 50 and 20 km/s clouds could be within a radius of 30 pc from Sgr A*. However, the \cite{Tress20} simulations do not include the gravitational field of the nuclear stellar disk, nuclear stellar cluster, or Sgr A* in their adopted potential  \citep[for reviews of these structures, see e.g. ][]{Launhardt02,Schoedel20,Schultheis21,Henshaw23}. As a result, the orbital kinematics of the gas in the inner parts of snapshots from their model (see their Figure 24, which they compare to the CMZ), including the inner 50 pc disk, are not reliable, and do not match well the observed constraints from the R$\sim$2-5 pc CND. A more realistic potential should lead the modeled velocity gradient of gas at these distances to be steeper, and thus inconsistent with the observational constraints from the 50 and 20 km/s cloud. Thus while these clouds could indeed lie on an orbital spur, the limits from the \cite{Kruijssen15} work still do not allow the gas to be as close as 10 pc from Sgr A*, and the \cite{Tress20} results should not be interpreted as compatible with that model.  

The \cite{Tress20} simulations also illustrate another potential contribution to the steeper velocity gradient observed across the 20 and 50 km/s clouds. These simulations show substantial velocity structure along orbital streams, with variations in the velocity gradient of the gas forming quasi-periodic `corrugations' in the orbital streams. Corrugations with wavelengths of $\sim$ 20 parsecs and amplitudes of $\sim$3.7 km/s have also been directly observed in CMZ gas \citep{Henshaw16b}. This wavelength, which \citeauthor{Henshaw16b} note is similar to the predicted Toomre length of 17 pc for CMZ gas, is comparable to the longitudinal extent of the 50 and 20 km/s clouds, as can be seen in Figure \ref{fig:Fig-PV}. If the same corrugations were superposed on the orbit of these clouds, they could then increase the overall change in velocity over a distance equal to half of the corrugation wavelength (or 10 pc) by 7.4 km/s. The change in velocity of the \cite{Kruijssen15} orbit at this location over a distance of 10 pc is $\sim$12 km/s (a velocity gradient of 1.2 km s$^{-1}$ pc$^{-1}$). The observed velocity change of the clouds over the same distance is 33 km/s (a difference of 21 km/s from the \cite{Kruijssen15} orbit, for a velocity gradient of 3.3 km/s). A corrugation identical to that observed by \cite{Henshaw16b} is then not enough to fully explain the steeper velocity gradient between these clouds, but could still contribute to their observed velocity structure. Alternatively, another possible contribution to the steepening of this velocity gradient could be the acceleration of the cloud from its interaction with the expanding Sgr A East SNR \citep{Genzel90}.

A more recent model by \cite{Sofue25} attempts to explain the kinematics of CMZ gas with a six-arm spiral model, in which the gas is moving on circular orbits (with fixed orbital velocity) at various radii. They suggest that the 20 km/s cloud is located on `Arm III' at a radius of 42 pc, while the 50 km/s cloud is located closer to Sgr A* on `Arm V' at radius of 8 pc. Our observations do not support this interpretation; the velocity gradient of the 50 km/s cloud appears flat (rather than steep, as required by such a small orbital radius). We do not see an obvious connection to the faint feature in Figure \ref{fig:Fig-PV} at a position of 6.0 pc between velocities of -50 and -100 km/s which \cite{Sofue25} suggest makes up the rest of `Arm V'. The 50 km/s cloud instead appears smoothly connected to the molecular ridge and the 20 km/s cloud. While \cite{Sofue25} argue that the larger linewidths of the 50 km/s cloud are not consistent with the molecular ridge and 20 km/s cloud, we suggest that this is a natural result of its interaction with the nearby SNR Sgr A East, and should not be interpreted as evidence for its being kinematically unrelated to the 20 km/s cloud. 

In summary, the kinematics strongly indicate that apart from the CND, none of the other molecular gas observed toward the central 10 pc is actually orbiting as close to Sgr A* as it appears in projection. Thus, while it is still possible (and perhaps likely, given their velocity gradient) that the 50 and 20 km/s clouds do not lie on a shared orbit with other CMZ gas \citep[as argued by][and Lipman et al. submitted]{Tress20}, the motions of these clouds cannot be reconciled with lying as close to Sgr A* as the \cite{Herrnstein05} model requires. Allowing either clouds to be located within 5 pc of Sgr A* would require these clouds to be undergoing highly unusual motions, for which no compelling examples have been presented from existing models or simulations.

\subsection{Revisiting Radio Absorption Measurements}
\label{sec:synchrotron}

\begin{figure*}[tbh]
\includegraphics[width=1.0\textwidth]{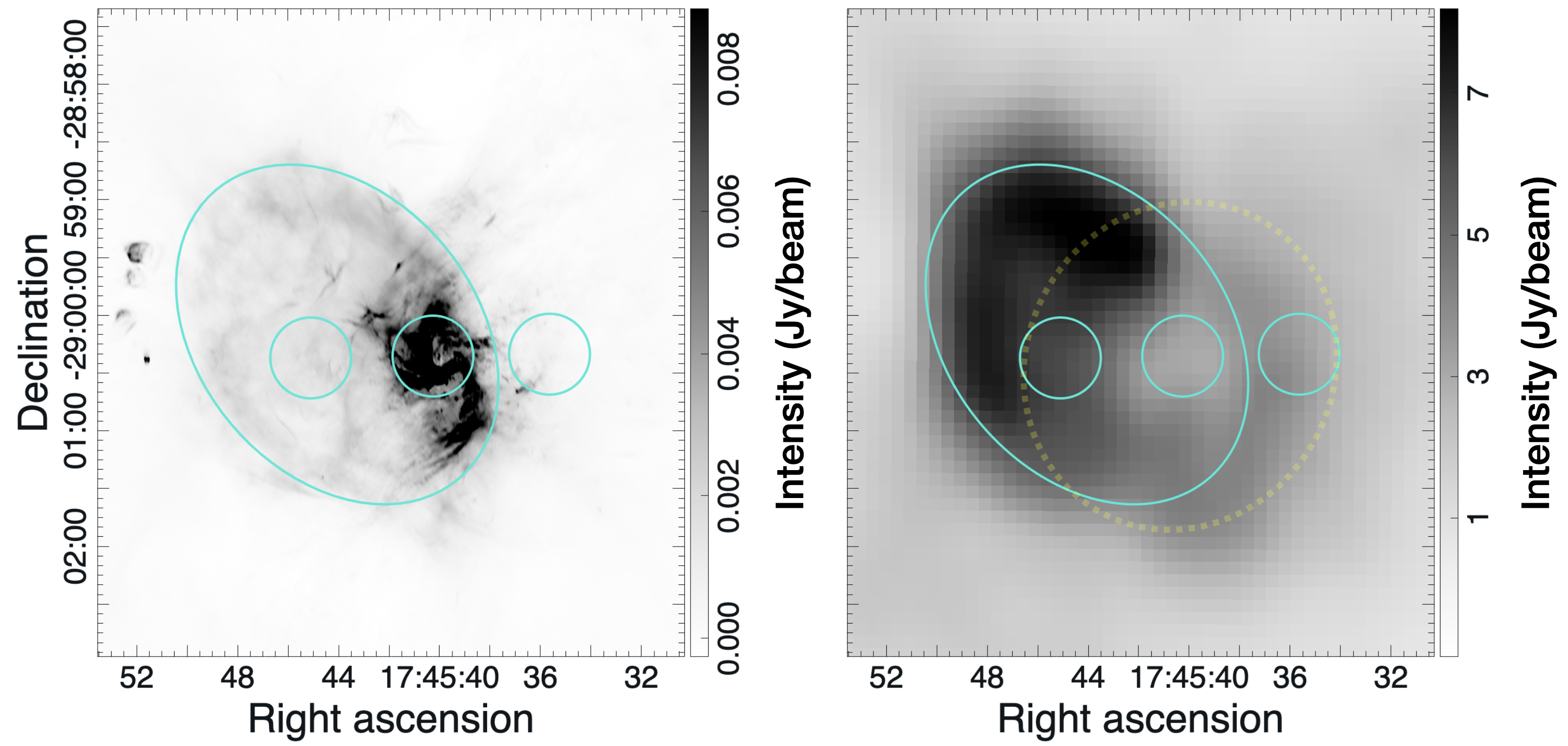}
\caption{Left: A 6 cm radio map of emission toward Sgr A East and West from \cite{Zhao16}. The {cyan} ellipse marks the shape and extent of the Sgr A East supernova remnant shell at this wavelength.  Right: A 90 cm map of the same region from \cite{Pedlar89}. Sgr A East appears in emission, while emission toward Sgr A West is depressed, appearing as an absorption feature. The three {cyan} apertures toward the center of Sgr A East, in the minispiral, and to the west of both the minispiral and Sgr A East are used to extract fluxes from the 90 cm image. {The dashed light yellow ellipse marks an approximate extent of a hypothesized synchrotron background component centered on Sgr A* discussed in Section \ref{sec:synchrotron}.}  }
\label{fig:Fig-90cm-measure}
\end{figure*} 

We now revisit the major constraints for the \citetalias{Herrnstein05} model. As discussed in Section \ref{intro}, a central assumption of models placing the 50 and 20 km/s clouds in close proximity to Sgr A* is the requirement that Sgr A East lies behind Sgr A West. This derives from 90 cm continuum observations that appear to show the Sgr A West minispiral as an absorption feature against the Sgr A East SNR shell \citep{YusefZadeh87,Pedlar89}. This constraint, combined with strong evidence for the interaction of the 50 km/s cloud with Sgr A East, strongly limits the placement of the 50 and 20 km/s clouds. 

A supposition in ascribing the 90 cm structure of Sgr A East to absorption by the Sgr A West minispiral is that the Sgr A East shell is uniformly limb-brightened along its edge. In this case, the lack of strong emission at 90 cm from Sgr A East on its western edge is then a result of all of the emission from the edge of the Sgr A East shell in this region being absorbed by Sgr A West. However, in Figure \ref{fig:Fig-90cm-measure}, it can be seen that there is also a reduction in emission in the southeast part of the shell, which is not due to the location of the minispiral or other thermal emission. We instead suggest that the shell is intrinsically brighter on its northeast side than on its southwest side. Specifically, we hypothesize that the interaction with the 50 km/s cloud to the northeast is the cause of this non-uniform limb brightening in the Sgr A East SNR. Similar signatures are seen in other supernova remnants that are hypothesized to be interacting with molecular clouds \citep[see e.g., G5.4-1.2, G349.7+0.2, IC 443;][]{Hewitt09,Lazendic10,Castelletti11}. While the steep reduction in brightness of the Sgr A East shell on its northwest edge does appear close to where it overlaps with Sgr A West, we propose that this is a coincidence of alignment, noting as well that the location where the shell brightness initially drops is somewhat offset from the position of the northern arm of the minispiral. 

Assuming that the bright edge emission observed in the northeast part of Sgr A East is not representative of the brightness of the shell where it overlaps with Sgr A West, we place three representative apertures (the size of the 90 cm beam) on the 90 cm image. These apertures are shown in Figure \ref{fig:Fig-90cm-measure}. Aperture 1 is placed away from the shell edge, toward the center of Sgr A East. Aperture 2 is placed toward the center of the minispiral, covering the northern and eastern arms. Aperture 3 is placed to the west of both Sgr A East and the minispiral. For Aperture 1, we measure a mean intensity of 6.13 Jy. For Aperture 2, we measure a mean intensity of 3.27 Jy. For Aperture 3, we measure a mean intensity of 3.44 Jy. The first thing we note is that the 90 cm emission toward the minispiral is non-zero. This is not entirely unexpected, as the entire volume of the CMZ is believed to permeated with a relatively high cosmic ray density \citep[cosmic ray ionization rates of $>$ 10$^{-15}$-10$^{-14}$ s$^{-1}$;][]{Goto14,Oka19,YusefZadeh13}. These cosmic rays are responsible for extended meter-wave synchrotron emission \citep[e.g.,][]{LaRosa05}. However, the relatively large amount of emission compared to the background level measured in identical apertures farther from Sgr A ($<$0.5 Jy) likely requires an additional source of emission. Some possibilities include: 

\begin{enumerate}
\item The minispiral is in front of Sgr A East and is absorbing all of the emission from Sgr A East. However, the minispiral itself only takes up a small fraction of the aperture (half or less), such that the emission and absorption (i.e., zero emission) average to an intermediate value. While this is a possibility, our comparison of these apertures with 6 cm emission (from an image with much higher spatial resolution) shows this is unlikely to be the case: the minispiral emission appears to take up the majority of this aperture. A spatially-extended region of positive spectral index can also be seen toward the minispiral in MeerKat images, e.g. Figure 14 of \cite{Heywood22}.  
\item The minispiral is inside of the Sgr A East shell, and so there is some emission from the shell in front of Sgr A West that is not absorbed. While this scenario has been suggested by \cite{Maeda02} and adopted in the \citetalias{Herrnstein05} model, we find that this is incompatible with the geometric constraints discussed previously, as the pericenter distance of the 50 (and 20 km/s) cloud would need to be much closer than 10 pc for Sgr A East to be both enveloping Sgr A West and interacting with the 50 km/s cloud. 
\item The minispiral is behind Sgr A East. In this case, all of the emission measured in the aperture toward Sgr A West would be representative of the intrinsic 90 cm brightness of the supernova remnant at that position, and could even have an additional contribution from (possibly optically-thick) free-free emission from Sgr A West. 
\end{enumerate}


As the minispiral shape of Sgr A West is clearly seen in silhouette rather than in emission, it must still lie in front of some source of background synchrotron emission. In this case, Sgr A West might be expected to absorb the entirety of any emission located behind it, and the brightness of Sgr A East at this position would appear diminished, as it would lack the additional emission signal added by its superposition with this background. This idea is supported by Figure 14 from \cite{Heywood22}, in which it can be seen that Sgr A East does not stand out in a map of spectral index derived from 20 cm MeerKAT observations. The emission surrounding Sgr A* out to radii greater than 3$'$ has a uniformly-negative spectral index indicative of synchrotron emission, with the two exceptions of the minispiral and the Sgr A HII regions. We note that the Sgr A HII regions, which do not overlap with the Sgr A East shell, do not appear in emission at 90 cm (Figure \ref{fig:Fig-90cm-measure}). This could either be because they are also unresolved absorption features against this background, or because they are sufficiently optically thick at these wavelengths so as not to be detectable.  


We {note that there is 90 cm emission surrounding the minispiral that is visible in Figure \ref{fig:Fig-90cm-measure}, some of which lies outside the observed bounds of Sgr A East. We mark an approximate extent of this emission with a light yellow ellipse with a radius $\sim$ 3 pc. We} suggest that this {could be a local synchrotron background} centered on Sgr A*, {which would} represent a local enhancement due to activity either from Sgr A* or, more likely, the nuclear cluster, which with at least one young population component having an age of 2.5 - {5.8} Myr \citep{Lu13}, {is likely to have experienced} recent supernovae. {Given the strength of the magnetic field in the Galactic center region, it is reasonable to ask whether the lifetime of synchrotron electrons from these sources is consistent with a persistent synchrotron halo. Field strengths of 1-5 mG observed toward the central parsec \citep{Killeen92,Plante95,Guerra23} correspond to a synchrotron lifetime of 10,000 - 100,000 years \citep{Condon92}. We assess whether supernovae alone could be frequent enough to maintain the synchrotron electron population by using pySTARBURST99 \citep{Hawcroft25} to estimate the average supernova rate for the young nuclear cluster. Following \cite{Lu13}, we assume an initial mass function with a slope of 1.7 and a total cluster mass in stars above 1 M$_\odot$ between 14,000 and 37,000 M$_\odot$. We also adopt a Galactic center metallicity of $Z=0.02$, though this does not strongly influence the resulting supernova rate. The average supernova rate for the time interval between 3 Myr (when the first supernova is estimated to occur, assuming an upper mass of 120 M$_\odot$) and 5.8 Myr ranges from 3.4$\times10^{-5}$ - $8.9\times10^{-5}$, depending on the adopted cluster mass. This corresponds to an average time between supernovae of 11,000-30,000 years, which is potentially sufficient to sustain the synchrotron population assuming the average magnetic field is not substantially stronger than 5 mG.} 

{If the nuclear cluster is responsible for this synchrotron background,} we might expect that the synchrotron distribution would be strongest toward the central cluster, and likely would extend {with diminishing strength} above and below the galactic plane, in a roughly bipolar structure following the magnetic field geometry \citep[similar to observed X-ray structures;][]{Morris03,Ponti19}. Sgr A West would only block about half of this background, as some of this synchrotron emission would lie in front of the minispiral. This still does require then that the Sgr A East SNR be overall intrinsically fainter than average on its western side compared to its eastern side. 

\begin{figure}[tbh]
\includegraphics[width=0.45\textwidth]{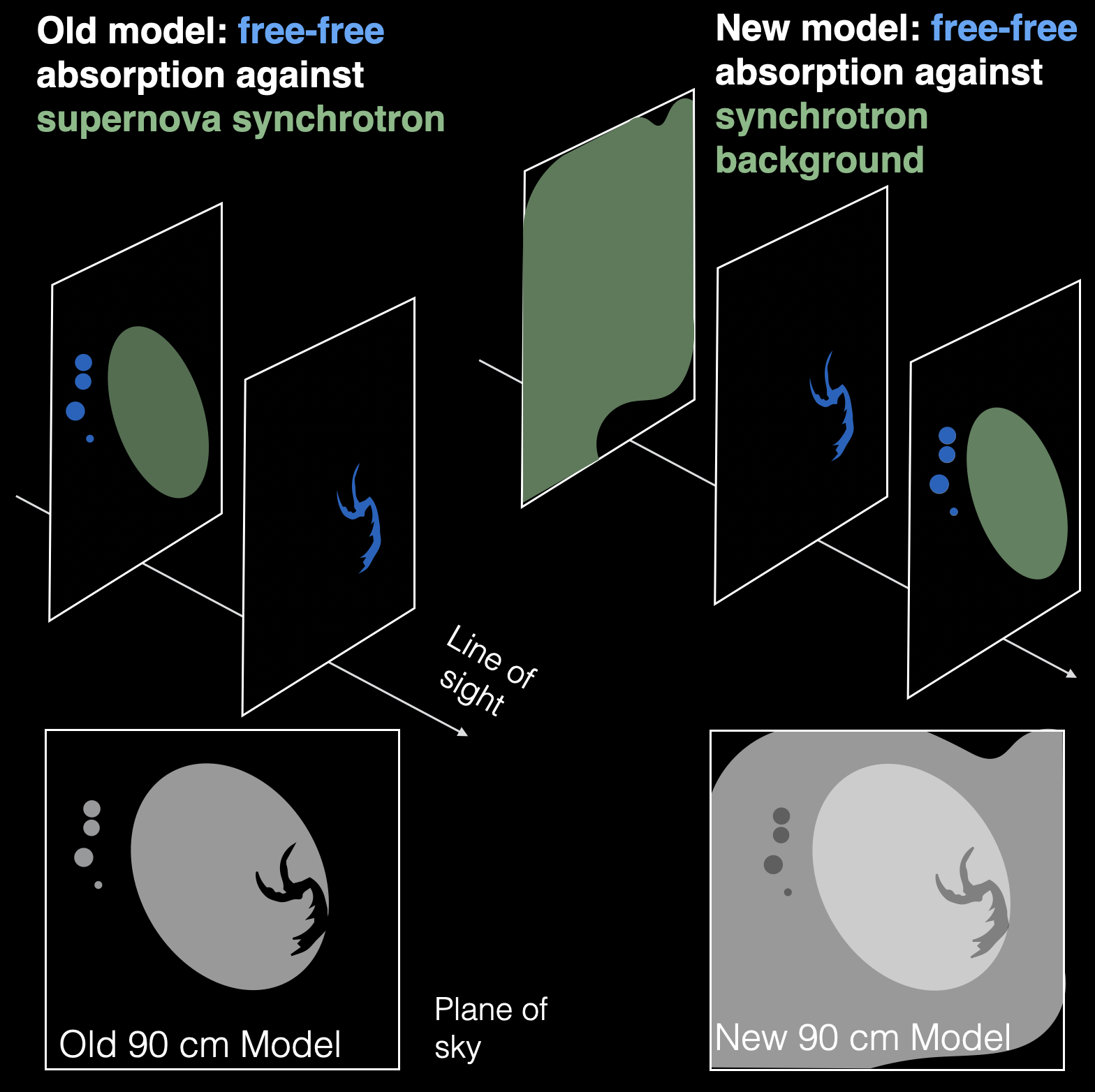}
\caption{A comparison of models for the observed 90 cm radio emission. At 90 cm, the free-free emission in the Sgr A West minispiral appears in absorption against a synchrotron background source. Existing models have Sgr A West appearing in absorption against the Sgr A East supernova remnant, requiring Sgr A West to be located in front of Sgr A East. However, here we suggest that the 90 cm observations could also be consistent with Sgr A West appearing in absorption against a Galactic center synchrotron background, which would allow Sgr A West to lie behind Sgr A East.} 
\label{fig:Fig-radio-model}
\end{figure}

In Figure \ref{fig:Fig-radio-model} we show a cartoon of the generally accepted model for the 90 cm morphology of Sgr A, and our proposed alternative. In the original model, Sgr A West is in front of the Sgr A East SNR, and absorbs the synchrotron emission from the SNR shell. In our alternative model, which explicitly calls for an additional background source of synchrotron emission, both the minispiral and the Sgr A HII regions would appear in absorption at 90 cm against the synchrotron background. Sgr A East is an additional source of synchrotron emission that is superposed on top of the emission and absorption from these other sources. Such a scenario should be testable with future, higher-resolution radio observations at low frequencies with the Square Kilometer Array. Such observations can more clearly show whether the brightness of the Sgr A East shell declines independently of any overlap with the minispiral, and can enable more robust measurements of the degree to which the minispiral exhibits absorption relative to Sgr A East and any more extended background emission. 

\subsection{An Updated Model of the Inner 10 Parsecs}

\begin{figure*}[tbh]
\includegraphics[width=1.0\textwidth]{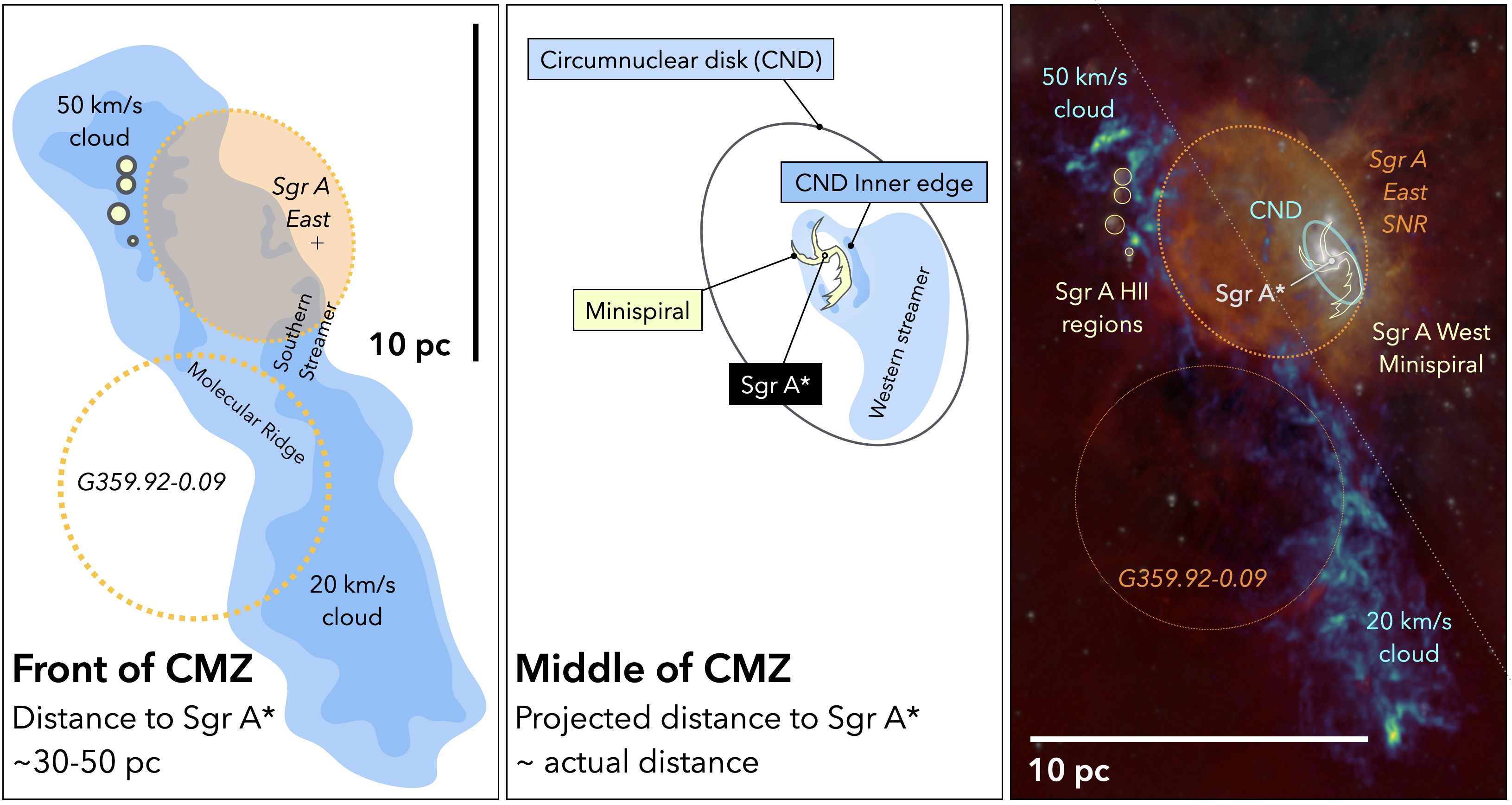}
\caption{An alternative model for interpreting observations toward the central 10 parsecs of the Milky Way. All images are shown in the plane of the sky, as seen by an observer. Structures observed along the line of sight toward the central 10 parsecs are separated into two groups. In the left panel, we show those located at some distance in the foreground (as discussed in the text, we prefer a distance of 30 - 50 pc, though this distance is not directly constrained by our new observations). In the middle panel, we show those located at the distance of Sgr A*. For the latter group, the projected separation to Sgr A* is roughly equivalent to their actual distance. In the right panel, for comparison, we again show a labeled view of multiwavelength observations from Figure \ref{fig:Fig-overview}. }
\label{fig:Fig-newmodel}
\end{figure*} 

\begin{figure}[tbh]
\includegraphics[width=0.49\textwidth]{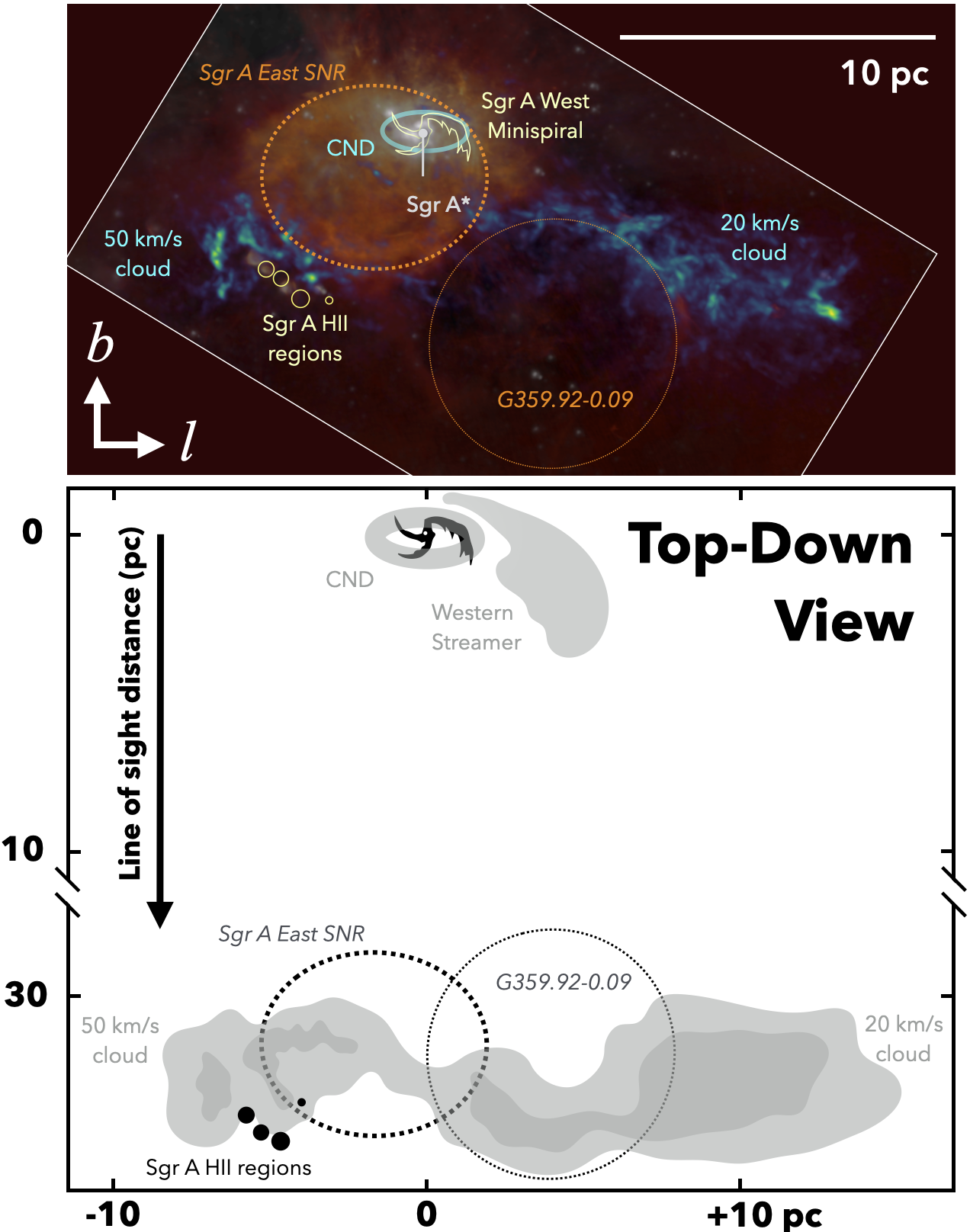}
\caption{Top: Multiwavelength overview of observations from Figure \ref{fig:Fig-overview}, rotated to be in Galactic coordinates. Bottom: A birds-eye-view of our alternative model for interpreting observations toward the central 10 parsecs of the Milky Way. The x-axis is distance in parsecs along the direction of Galactic longitude, while on the y-axis, we show the distance in parsecs along the line of sight toward Sgr A* (which is located at 0,0). In this diagram, we place the 50 and 20 km/s 30 pc in front of Sgr A*. However, as the exact distance is not well constrained, we draw the y-axis as a broken scale, which can also be interpreted as showing a minimum line-of-sight distance of 10 pc for these clouds. 
}
\label{fig:Fig-new-top-view}
\end{figure} 

Based upon the results of our analysis, we propose an alternative model for the location of structures in the inner 10 pc. Our main goal with this model is not to provide an exact placement of all of the structures in three dimensions relative to Sgr A*, but rather to make a general division between structures that are truly close to Sgr A*, and those located at larger distances. As shown in Figure \ref{fig:Fig-newmodel}, we propose that the structures identified in the \citetalias{Herrnstein05} model naturally fall into two groups. The first group, consisting of the Sgr A West minispiral, CND, and `Western Streamer', we locate in roughly the same plane of the sky as Sgr A* (at the same distance from the observer as Sgr A*). Only these sources are truly located within a radius of 10 pc of Sgr A*. The second group, consisting of the 50 and 20 km/s clouds (including the `Southern Streamer' {and} `Molecular Ridge') are all connected structures, and are located in front of Sgr A*, at a distance of more than 10 pc \citep[consistent with the detailed analysis of][]{Kruijssen15}, but more likely at least $\sim$ textbf{15 - 50} parsecs, consistent with recent constraints from infrared and radio absorption \citep[e.g.,][Lipman et al. (submitted)]{Walker25,Lipman25,NoguerasLara26}.  {We do not include the `Northern Ridge' as we cannot definitively associate it with either the 50 or 20 km s$^{-1}$ cloud.  We additionally} place the SNR Sgr A East, in this second group. The compressed ridge morphology of the 50 km/s cloud \cite{Serabyn92} and the presence of 1720 MHz OH masers at the velocity of the gas in this cloud \citep{YusefZadeh96,YusefZadeh03,Sjouwerman08} establish that Sgr A East is interacting with the 50 km/s cloud.

In contrast, the nature of G359.92-0.09 and its association with the `Molecular Ridge' and 20 km/s cloud is  more ambiguous. While bright 1720 MHz OH masers are also located in the gas at the northern perimeter of both this source and Sgr A East (in the `Molecular Ridge'), there is no report of masers located in the 20 km/s cloud along its western edge \citep[though this may be because it falls outside of the well-surveyed region around Sgr A East, see, e.g., ][]{YusefZadeh96,Karlsson03, Sjouwerman08}.  Overall, G359.92-0.09 is less studied than Sgr A East. It is a fainter source, more symmetric in appearance than Sgr A East in radio continuum images, though lacking continuous emission from its circumference, particularly on its southeast edge. Unlike Sgr A East (a mixed-morphology SNR) it does not have an X-ray bright core. It is most notable for the bundle of nonthermal filaments on the southwest edge of this source \citep[Sgr A-E; ][]{Ho85,Zhao16} which is also associated with an X-ray-emitting filament \citep{Sakano03, Lu03, Zhang14}. The classification of G359.92 as a supernova remnant is not definitive, and \cite{Zhang14} argue that the X-ray emission from the filaments on the southwest edge are not consistent with the interaction between a molecular cloud and a SNR. There are also arguments that this source represents part of a nuclear outflow from the the vicinity of Sgr A* and the young nuclear cluster. We tentatively associate this source with the 20 km/s cloud, following \cite{Hsieh16}, who argue that the morphology of the 20 km/s cloud and `Molecular Ridge' are potentially consistent with an interaction with G359.92-0.09. However, we acknowledge that the evidence for this association is still ambiguous. 


Our model is similar to the model originally proposed by \cite{Zylka90}. This model was particularly notable for considering orbital motion and attempting to reconcile structures on both 10 pc and 100 pc scales. The primary difference is that our model does not require Sgr A East to be behind Sgr A West. In the \cite{Zylka90} model this resulted in the complicated solution that the main body of the 50 km/s cloud (or `East Core'\footnote{The nomenclature in this paper requires some explanation, as \cite{Zylka90} also refer to a 50 km/s cloud, but this not the same as our 50 km/s cloud  (M-0.02-0.07; which they refer to as the `East core'). The 50 km/s cloud in their paper is a different cloud they suggest lies behind Sgr A*.}) was not connected to the `Molecular Ridge' of the \citetalias{Herrnstein05} model.  

Our model preserves the following aspects of the \citetalias{Herrnstein05} model: 

\begin{enumerate}
    \item Sgr A East and the 50 km/s cloud are interacting. 
    \item The 20 km/s cloud and 50 km/s cloud are connected by the `Molecular Ridge'. 
    \item The CND is connected to the minispiral.
    \item The 20 km/s cloud is in front of Sgr A*
\end{enumerate}    

It also results in the following changes to the \citetalias{Herrnstein05} model: 

\begin{enumerate}    
    \item The 20 and 50 km/s clouds are not connected to the CND
    \item The 20 and 50 km/s clouds are part of a common orbital structure (consistent with the finding of \citeauthor{Kruijssen15} who placed them on `Stream 1'), the kinematics of which are consistent with a location outside of the central 10 pc, given the relatively shallow velocity gradient across the clouds 
    \item The 20 and 50 km/s clouds are both in front of Sgr A*
    \item The `Western Streamer' is an outer part of the CND that is not interacting with Sgr A East or connected to the 20 km/s cloud.
\end{enumerate}

\subsection{Implications of the Alternative Model}

Below, we outline some important implications of this model.

\subsubsection{The Nature and Age of Sgr A East}

Our new location for Sgr A East, outside of the central 10 pc, has several impacts on our understanding of its origin and evolution. Modeling by \cite{Uchida98} has previously ascribed the slightly elongated shape of Sgr A East to shear due to its small orbital radius, though more recent modeling by \cite{Ehlerova22} suggests that shear alone may not significantly alter the shape on timescales less than the largest adopted age for Sgr A East ($10^4$ years). As tidal shear should be significantly less important to the evolution of Sgr A East if it is located at a larger distance from Sgr A*, its shape is likely solely due to expansion into an inhomogeneous medium, consistent with modeling by \cite{Ehlerova22}. This could indicate an origin near the edge of the 50 km/s cloud, similar to the placement of the Sgr A HII regions on the front side of the 50 km/s cloud \citep{Mills11}. 

Placing Sgr A East at a larger distance from Sgr A* also impacts the adopted age of this source. Younger ages for Sgr A East (e.g., on the order of $10^3$ years) come from models that assume that the supernova explosion occurred in a region of lower mean density within 2 pc of Sgr A*, and that it has interacted with stellar winds from the young nuclear cluster \citep{Rockefeller05,Yalinewich17,Zhang23}. These and other models \citep[e.g.; ][]{Uchida98,Rimoldi15} also use its location to put upper limits on age, assuming it would be strongly tidally distorted and become no longer visible after $<10^4$ years of evolution in close proximity to Sgr A*. By changing the environment of Sgr A East, our model thus supports an older age of at least $10^4 $~years, similar to that adopted by \cite{Ehlerova22}. 

Finally, while Sgr A East is generally accepted to be a relatively compact example of a mixed-morphology supernova remnant \citep{Maeda02}, this location does eliminate some of the more exotic suggestions for its origin \citep[e.g., as the remnant of a tidal disruption event from Sgr A*; ][]{Khokhlov96, Guillochon16}.

\subsubsection{Inflow to Sgr A*}
\label{inflow}

With the 50 and 20 km/s clouds both located outside of the central 10 pc, this removes a potential large reservoir of future accretion for the central supermassive black hole \citep[the combined mass of the 20 and 50 km/s cloud complex is estimated to be $\sim5.5\times10^5$ M$_\odot$; ][]{Battersby25b}. The CND is then the only major molecular gas reservoir in the central 10 parsecs and the mass of this reservoir \citep[$10^4$ M$_\odot$; ][]{RequenaTorres12} is less than 0.1\% of the total CMZ mass \citep[$\sim5\times10^7$ M$_\odot$ ; ][]{Dahmen98}, the majority of which is likely at radii of $50-100$ pc \citep{Kruijssen15}. This is part of an overall steep fall-off in the amount of gas mass available for accretion as a function of radius from Sgr A*: the estimated mass of neutral gas in the central parsec is a few hundred solar masses \citep{Jackson93}, while the mass of the ionized gas streamers in the minispiral at distances down to 0.17 pc is only a few tens of solar masses \citep{Zhao09}.

While we find that the 50 and 20 km/s clouds are located outside of the central 10 pc, this does not entirely prohibit connections between these clouds and gas associated with the CND in the central 10 pc. Indeed, if these clouds are close to a pericenter passage along their orbit, as argued by \cite{Kruijssen15}, they could be undergoing some tidal stripping, bringing gas from these clouds to smaller radii. However, our analysis of new \am data (see Section \ref{sec:Ammonia}) rules out all three features that were suggested to link the CND with these clouds in the \citetalias{Herrnstein05} model: from the 50 km/s to the CND (via the `Northern Ridge'), and from the 20 km/s cloud to the CND (via the `Southern Streamer' and `Western Streamer'). Our position-velocity data (Figure \ref{fig:Fig-PV}) are also inconsistent with simulations of \cite{Ballone19}, which are used to suggest that the 20 km/s cloud interacted with Sgr A* to form the CND. Our observations do not show the large spread of high-velocity material (v$>$50 km/s) that is seen in a simulated longitude-velocity diagram from \cite{Ballone19} Figure 4, and which would represent a kinematic link between the 20 km/s cloud and the CND. Instead, we observe the 20 km/s cloud as a narrow-width ($\sim$ 5 km/s) feature with a shallow velocity gradient (2.5 km s$^{-1}$ pc$^{-1}$) that continues without disruption to the greater longitudes of the 50 km/s cloud. 

We assert that remaining evidence for additional connections, for example as suggested by \cite{Takekawa17}, is not yet compelling, given the relatively low spatial resolution of the data used to make these claims and the large amount of spatial overlap between the orbital structures of the CND and the 50 and 20 km/s clouds. One should expect a significant amount of superposition among the distinct orbital components located along the line of sight to the Galactic center, given our edge-on view of this region. With such complex velocity structures in this region \citep[beyond just orbital motion, there are many unusual velocity features in the CMZ, variously attributed to expanding shells, SNRs, intermediate mass black holes, and colliding clouds in CMZ gas, e.g.; ][]{Oka11,Oka16,Butterfield18,Yalinewich18,Takekawa19,Gramze23,Ginsburg24} there must be a higher threshold of evidence for true interaction than finding extended structures that overlap in position-velocity space.

Without clear evidence for a connection between {the} 50 and 20 km/s {clouds and the CND}, this implies that the CND may not be a persistent or long-lived feature, though there are likely other avenues for inflow of less dense gas onto this source \citep[e.g.][]{Tress24}.  The total amount of gas observed to currently exist in the central cavity interior to the CND is relatively small \citep[a few hundred solar masses, less than 3\% of the entire CND mass;][]{Jackson93,Zhao09}. While some interpretations of CND kinematics claim to detect an inflow component to its overall motion, with inflow representing up to 25\% of the overall gas motion \citep{Wardle90,Oka11,Hsieh18}, these analyses interpret all non-circular motion as pure inflow. In contrast, other models suggest that inflow proceeds more slowly. \cite{Sato24} suggest an inflow rate of $10^{-2}$ M$_\odot$ per year, while \cite{Solanki23} base a much lower estimate on observations of a small (M$\sim10^{-4}$ M$_\odot$) accretion disk of warm gas around Sgr A* \citep{Murchikova19}, which they suggest forms over thousands of years due to the result of interaction with the winds from the central star cluster. In contrast to these treatments of gas inflow, it should be noted that measurements of gas density that indicate a lack of gravitational stability of individual clumps in the CND \citep{RequenaTorres12,Mills13,Hsieh21} only require that the current substructure observed within the CND is transient, not that the entire structure is short-lived \citep{Blank16}. While some amount of feeding is likely occurring via extensions of the `Western Streamer' observed in \am \citep{Hsieh17}, it is still not clear how or whether this connects to the main reservoir of CMZ gas. 

As mentioned in the previous section, an important difference of our model for studies of the central parsec is that Sgr A East does not overlap with or enclose Sgr A West, the ionized gas minispiral that surrounds Sgr A*. Many previous hydrodynamic models of the central few parsecs assumed that the Sgr A East SNR was interacting with the stellar winds in the central light year, though they did not agree on whether the passage of the shell would clear the central 0.2 pc of gas, limiting accretion onto Sgr A* \cite{Maeda02}, or have no effect on this region \citep{Rockefeller05,Rimoldi15}. However, with Sgr A East located outside of the central 10 parsecs, the magnitude of the accretion flow onto Sgr A* \citep[including evidence of flares from Sgr A* over the past few hundred years; ][]{Sunyaev98,Ponti10}, must be understood without invoking interaction with this SNR. X-ray features like the `ridge' \citep{Baganoff03} that are interpreted as evidence of interaction between Sgr A East and the central parsecs would also require an alternative explanation. 


\subsubsection{Molecular Cloud Lifetimes in the CMZ}

The morphology and size of Sgr A East suggest it occurred in a relatively high-density environment associated with the 50 km/s cloud \citep{Maeda02,Ehlerova22}. If the progenitor of Sgr A East formed in the 50 km/s cloud, this would imply that at least a substantial part of this cloud has persisted in the CMZ for more than a few million years after the initiation of star formation \citep[e.g., 3 million years for the most massive stars; ][]{Schaller92,Zapartas17}. This is consistent with estimates of CMZ cloud lifetimes\citep[1.4-3.9 Myr; ][]{Jeffreson18}, and similar to the estimated orbital period for CMZ gas  \citep[2-3 Myr; ][]{Kruijssen15}. The 50 km/s cloud does also contain the Sgr A HII regions \citep{Mills11,Sjouwerman96}, which while they must be at least $\sim 10^4$ years old, could be significantly older. 

However, there is also good reason, due to the different forces they are subject to, to expect any formed stars to orbit differently than the gas on $\sim$ 5-10 Myr timescales. This is true even though overall there is good agreement in the kinematics between the gas and the nuclear stellar disk \citep{Schultheis21,NoguerasLara24,Schultheis25}. An alternative to in-situ formation is that the progenitor of Sgr A East was an interloper, for example a tidally stripped member of another young cluster or nearby association \citep{Habibi14}. It should be noted that while the Sgr A HII regions are generally accepted to have been formed in-situ in the 50 km/s cloud \citep{YusefZadeh10,Mills11}, it has also been suggested that the ionizing stars for these regions could be interlopers as well \citep{Lau14}. 

One can also ask, given the star formation that has already occurred in the 50 km/s cloud and the current impact of Sgr A East, how much longer will this cloud exist? From the analysis presented in \cite{Battersby25b}, the 50 km/s cloud is physically smaller than the 20 km/s cloud (an area of 22 pc$^2$ compared to 78 pc$^2$) and substantially less massive ($6.2\times10^4$ M$_\odot$ compared to $3.1\times10^5$).  On average the 50 km/s cloud is estimated to be somewhat more dense ($1.2\times10^4$ cm$^{-3}$ compared to $8.6\times10^3$ cm$^{-3}$), however it has a lower peak column density ($1.8\times10^{23}$ compared to $4.2\times10^{23}$). There is less dense substructure in the 50 km/s cloud compared to the 20 km/s cloud \citep[][; see their Figure 10]{Battersby20}. Unlike the 20 km/s cloud, in which there are numerous water masers indicative of ongoing star formation \citep{Guesten83,Sjouwerman02,Lu15}, the 50 km/s cloud also does not show clear evidence of very recent stages of star formation.  Given an age of $10^4$ years for years for Sgr A East, this absence of continuing star formation could plausibly be attributed to supernova feedback. The recent impact of the Sgr A East SNR is at least having a clear impact on the chemistry of the 50 km/s cloud \citep{Tanaka11,Tanaka21}. Interestingly, if G359.92-0.09 is a SNR interacting with the 20 km/s cloud, it is having a much more subdued impact on both the cloud chemistry and ongoing star formation. Given the lack of ongoing star formation in the 50 km/s cloud, one could argue that this is an example of a CMZ molecular cloud reaching the end of its lifetime. Further, if the progenitors of Sgr A East and G359.92-0.09 are not associated with the 50 and 20 km/s clouds, it suggests that disruption due to interloper stars is an important effect that should be considered in theories of cloud lifetime in this environment. 




\section{Conclusions}
\label{conclusions}

We have presented a suite of new observations toward the central 10 pc of the Milky Way, including a sensitive new \am map of the inner $11 \times 22$ parsecs, ALMA maps of the CND in multiple molecular lines, and a \co 1-0 absorption spectrum toward Sgr A*. Based on our analysis of these data, we have developed an alternative model for the location of structures in the central 10 pc of our galaxy in which the 50 and 20 km/s clouds and Sgr A East SNR lie more than 10 pc \citep[{but more likely at least 15 - 50 pc, based on results from}][and Lipman et al. submitted]{NoguerasLara26} in front of Sgr A*, while the central 10 parsecs contains just the CND and Sgr A West. To support this model, we present an alternative explanation for the observed 90 cm morphology of Sgr A, which allows Sgr A West to be located behind Sgr A East. In this way, we are able to create a model of the central 10 pc that is consistent with the existing observational constraints of this region, and can be reconciled with recent orbital fits to gas kinematics of the entire CMZ. 

Overall, our analysis illustrates the complexities of interpreting a crowded region with many components along the line of sight that may exhibit chance alignments. While we are able to place some limits on the locations of the 50 and 20 km/s clouds, their exact distance along the line of sight still remains difficult to determine. We suggest a careful comparison of future kinematic models and simulations with improved data sets as well as careful modeling of cloud extinction and stellar distribution with facilities like JWST is likely the best way to obtain a more accurate line of sight distance to these structures. 

\section{Acknowledgments}
{We thank the referee of this paper for their constructive comments that strengthened our analysis and its presentation in this paper.}
E.A.C.\ Mills  gratefully  acknowledges  funding  from the National  Science  Foundation  under  Award  Nos. 1813765, 2115428, 2206509, and CAREER 2339670, and through the SOFIA archival research program under Award No.  09$\_$0205. 
C.\ Battersby  gratefully  acknowledges  funding  from  National  Science  Foundation  under  Award  Nos. 2108938, 2206510, and CAREER 2145689, as well as from the National Aeronautics and Space Administration through the Astrophysics Data Analysis Program under Award ``3-D MC: Mapping Circumnuclear Molecular Clouds from X-ray to Radio,” Grant No. 80NSSC22K1125.
A.\ Ginsburg acknowledges support from the NSF under AAG 2206511 and CAREER 2142300.
MCS acknowledges financial support from the European Research Council under the ERC Starting Grant ``GalFlow'' (grant 101116226) and from Fondazione Cariplo under the grant ERC attrattivit\`{a} n. 2023-3014.
FNL gratefully acknowledges financial support from grant PID2024-162148NA-I00, funded by MCIN/AEI/10.13039/501100011033 and the European Regional Development Fund (ERDF) “A way of making Europe”, from the Ramón y Cajal programme (RYC2023-044924-I) funded by MCIN/AEI/10.13039/501100011033 and FSE+, and from the Severo Ochoa grant CEX2021-001131-S, funded by MCIN/AEI/10.13039/501100011033.

\bibliography{central10pc}{}
\bibliographystyle{aasjournalv7}

\software{
          Miriad \citep{Sault95},
          }

\end{document}